\documentclass[9pt]{elife}

\title{On the potential of machine learning to examine the relationship between sequence, structure, dynamics and function of intrinsically disordered proteins}

\author[1,*]{Kresten Lindorff-Larsen}
\author[1,*]{Birthe B. Kragelund}
\affil[1]{Structural Biology and NMR Laboratory \& Linderstr{\o}m-Lang Centre for Protein Science,
Department of Biology, University of Copenhagen.  Ole Maaløes Vej 5, DK-2200 Copenhagen N, Denmark}

\corr{lindorff@bio.ku.dk}{KLL}
\corr{bbk@bio.ku.dk}{BBK}

\begin{document}

\maketitle

\begin{figure}[h]
\centering
\includegraphics[width=1.0\linewidth]{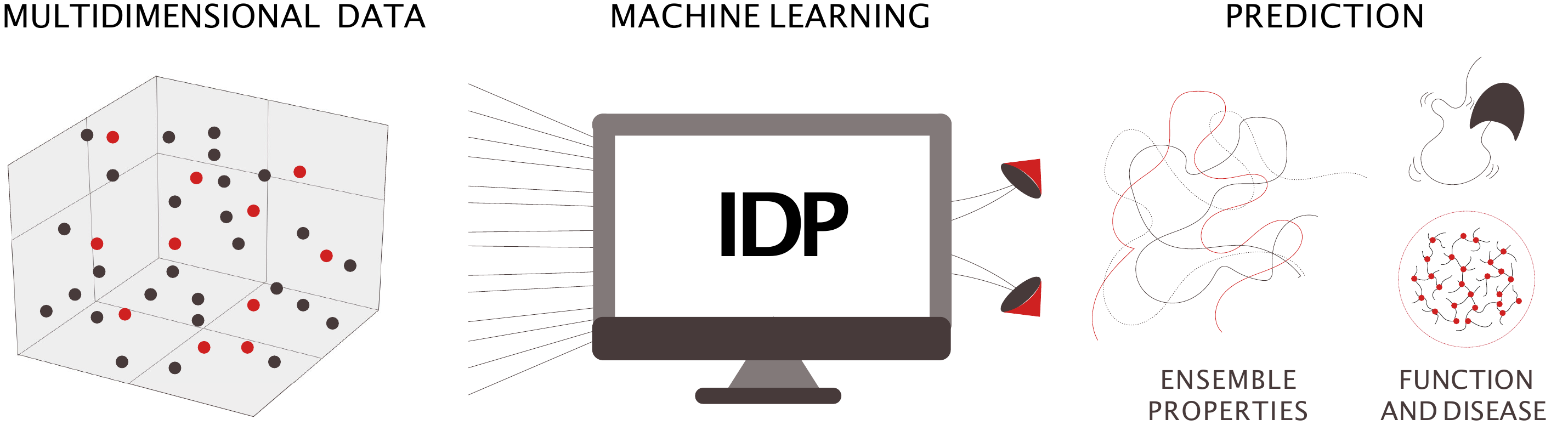}
\end{figure}

\begin{abstract} 
Intrinsically disordered proteins (IDPs) constitute a broad set of proteins with few uniting and many diverging properties. IDPs---and intrinsically disordered regions (IDRs) interspersed  between folded domains---are generally characterized as having no persistent tertiary structure; instead they interconvert between a large number of different and often expanded structures. IDPs and IDRs are involved in an enormously wide range of biological functions and reveal novel mechanisms of interactions, and while they defy the common structure-function paradigm of folded proteins, their structural preferences and dynamics are important for their function. We here discuss open questions in the field of IDPs and IDRs, focusing on areas where machine learning and other computational methods play a role. We discuss computational methods aimed to predict transiently formed local and long-range structure, including methods for integrative structural biology. We discuss the many different ways in which IDPs and IDRs can bind to other molecules, both via short linear motifs, as well as in the formation of larger dynamic complexes such as biomolecular condensates. We discuss how experiments are providing insight into such complexes and may enable more accurate predictions. Finally, we discuss the role of IDPs in disease and how new methods are needed to interpret the mechanistic effects of genomic variants in IDPs. 
\end{abstract}

\noindent{Keywords}: Machine learning, intrinsically disordered protein, molecular complex, condensate, SLiM

\section{Highlights}
\begin{itemize}
    \item Machine learning may help extract conformational preferences from poorly defined multiple sequence alignments of intrinsically disordered proteins
    \item Machine learning may help link biophysical experiments and conformational ensembles through better forward models
    \item Machine learning may help develop and parameterize better coarse-grained models for intrinsically disordered proteins
    \item Machine learning may help identify short-linear motifs, assign potential biological functions to them, and enable transfer of information between different systems
    \item Machine learning may help extract information from experiments on biomolecular condensates and enable transfer to proteomewide predictions
    \item Machine learning may help assign pathogenicity to genetic variants and be combined with other models to provide information about mechanisms of disease
\end{itemize}

\section{Introduction}
Intrinsically disordered proteins (IDPs) constitute a broad and relatively heterogeneous class of proteins that have in common that they do not adopt a well-defined three-dimensional structure, at least in the absence of binding partners. This in itself is not a very strict definition because also natively folded proteins are dynamic. Experimentally, disordered proteins are often characterized using a range of biophysical measurements that typically reveal the presence of transiently formed secondary structure elements and occasionally weak, transient longer-range interactions. Analogously, many proteins have regions of intrinsic disorder interspersed within or between folded domains, and in many ways these intrinsically disordered regions (IDRs) behave similarly to IDPs, and in general we will refer to both as IDPs.

The flexibility and dynamics combined with an extended surface area endow IDPs with an ability to adapt, a trait that is often key to their biological function, either because it enables them to bind to multiple different proteins or because the intrinsic dynamics may affect both binding kinetics and thermodynamics. This dynamics, however, also makes it difficult to characterize IDPs both experimentally and computationally.

It was early recognized that the amino acid composition and sequences of IDPs differed in several ways from those of folded proteins. Thus, aided by databases containing experimentally-validated IDPs \citep{hatos2020disprot} a large number of prediction methods have been developed to predict protein disorder from sequence alone \citep{piovesan2021mobidb,necci2021critical}. While overall very successful, such prediction methods inherently need to deal with the heterogeneity in what is considered a disordered protein, including large differences in biological context (complexes, post-translational modifications, etc).

A complementary approach to study IDPs is to characterize the conformational ensembles that they populate. In certain favourable cases, computational methods can on their own predict some conformational properties. Often, however, a detailed and accurate characterization requires integrating one or more types of biophysical experiments with computational methods to collectively derive a collection of structures that represent the conformational heterogeneity \citep{mittag2007atomic,jensen2013describing} or dynamics \citep{salvi2016multi} of the protein. A number of such approaches exist and some of the resulting ensembles are collected in the Protein Ensemble Database (PED; \cite{lazar2021ped}), by analogy to the Protein Data Bank (PDB; \cite{wwpdb2019protein}), which, however mostly contains more well-defined protein structures or IDPs in complexes.

For folded proteins, Anfinsen's observations \citep{anfinsen1973principles,eisenberg2018hard} suggested that it should be possible to predict the three-dimensional structure of a folded protein based only on its primary structure and its interaction with the environment. Over the years, this has led to the field of protein structure prediction, and a plethora of innovative approaches to predict structure from sequence. The accuracy of such methods are evaluated during the biennial critical assessment of structure prediction (CASP) experiment. While there have been continued improvements in the ability to predict structures over the years, the last two installments of CASP (CASP13 (2018) and CASP14 (2020)) have witnessed some substantial and impressive advances in accuracy, in particular in the so-called template-free modelling \citep{kryshtafovych2019critical,alquraishi2021machine}. While a number of developments have contributed to this, we here highlight three. First, in the last decade  a number of methods have been developed to extract structural information from multiple sequence alignments (MSAs) e.g. through the analysis of correlated substitutions during evolution \citep{lapedes2012using,weigt2009identification,marks2011protein,morcos2011direct,balakrishnan2011learning,xu2019distance}. Second, there has been an explosion in the number of sequences available making such sequence-based approaches useful and applicable to a wider number of proteins. Finally, various deep-learning approaches have been used to `learn' the complex relationship between the amino acid sequence (or MSA) and the three-dimensional structure. Most visible has been the development of the AlphaFold approach \citep{senior2020improved} in CASP13 and AlphaFold~2 in CASP14, although many other groups have also contributed to these developments including among others \cite{xu2019distance,zheng2019deep,kandathil2019recent,alquraishi2019end,torrisi2020deep, yang2020improved,alquraishi2021machine}.

\begin{figure}[tbp!]
\centering
\includegraphics[width=1.0\linewidth]{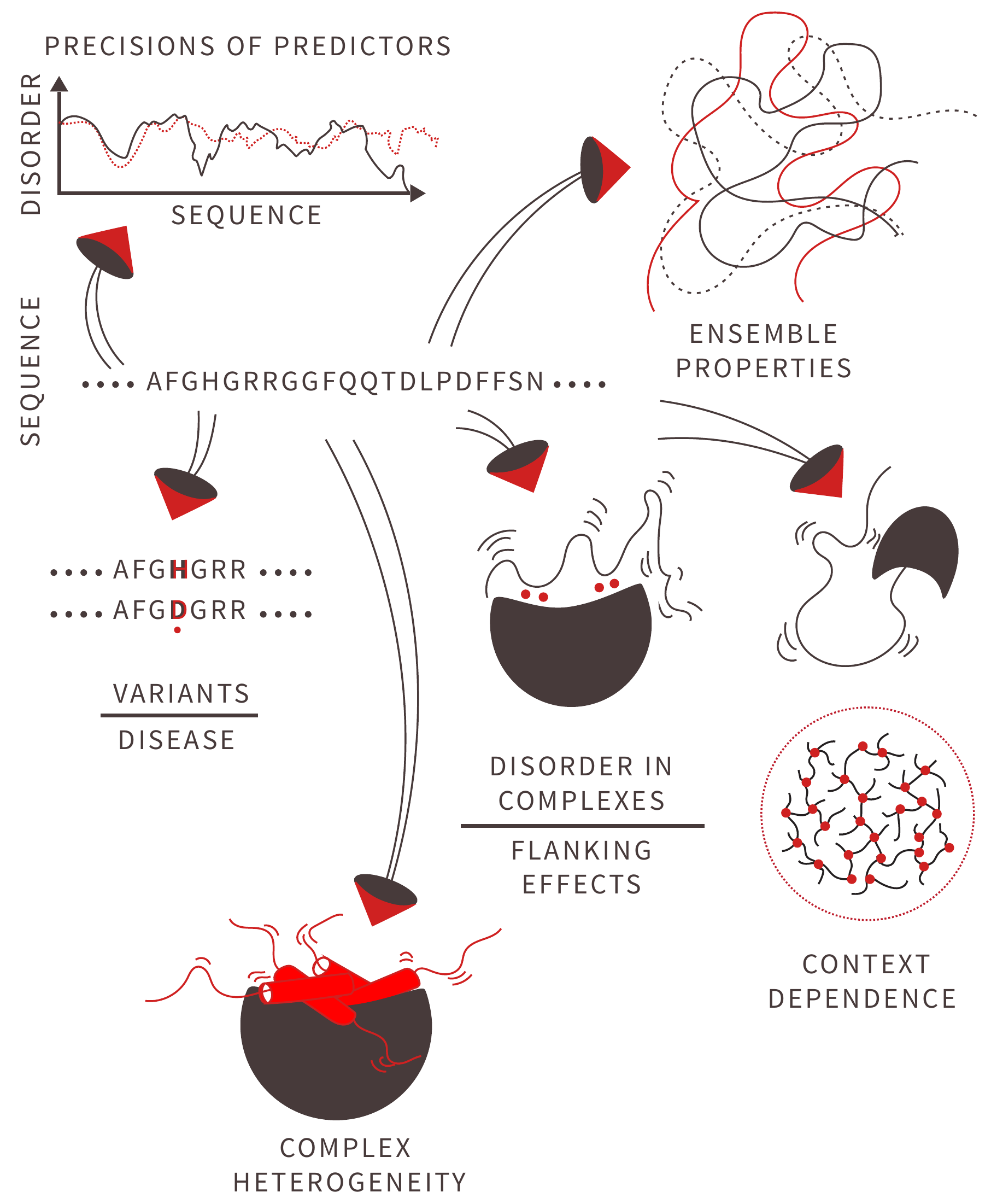}
\caption{An overview of the relationship between sequence, structure, dynamics and function of IDPs, and how the inherent disorder also in the complexes affects the use of machine learning approaches. Some of the challenges to understand these relationships include improving predictions of disorder, describing ensemble properties, and finding ways to include complex heterogeneity and context effects. Finally, it is still not clear how many disease variants in IDPs lead to disease and by which mechanisms.}
\label{fig:fig1}
\end{figure}

Motivated by our own research, this review begins by examining whether such methods can be used to predict information about the (highly conformationally heterogeneous) three-dimensional `structures' and ensembles of IDPs using only the primary structure as input. We also discuss how machine learning methods may aid in integrative modelling of the conformational ensembles of IDPs. We discuss the unique properties of IDPs in complexes, both those formed via short linear motifs and in larger assemblies and biomolecular condensates, and how new sources of data may be useful to develop better prediction methods. Finally, we discuss the role of IDPs in human diseases and how an improved understanding of the relationship between sequence, structural properties, formation of complexes and function may help in this area (Fig.~\ref{fig:fig1}). Overall, we highlight a number of challenges that are particularly relevant for IDPs, and some of the questions that might be addressed by combining machine learning methods with experiments and other computational approaches.

\section{Towards improved conformational ensembles}
\subsection{From sequence to structure}
Before discussing potential applications to IDPs, we first describe very briefly some of the key steps that have lead to improved structure prediction of folded proteins, but note that this description is far from comprehensive. One key ingredient has been the ability to extract structural information, for example in the form of contacts \citep{lapedes2012using,marks2011protein}, distance distributions \citep{senior2020improved} or distributions over distances and orientations \citep{yang2020improved}, from the analysis of MSAs. This work builds on earlier ideas that correlated mutations observed through evolution contain information about the proximity of amino acids in the three-dimensional structure \citep{taylor1994compensating,neher1994frequent,gobel1994correlated}, but required more advanced analysis methods, including global analysis methods \citep{lapedes2012using,weigt2009identification,marks2011protein,morcos2011direct,balakrishnan2011learning,xu2019distance}, as well as increased number of sequences to reveal their full potential. The structural information obtained from the MSAs can then be used to guide structure determination using a range of methods to obtain three-dimensional models. Indeed, while many methods have been shown to help improve the accuracy of contact prediction, it is not clear that improved contact prediction always leads to substantially improved models of three-dimensional structures \citep{kassem2018enhancing}. More recent improvements in the AlphaFold~2 method appear to involve using a so-called end-to-end model \citep{alquraishi2019end,laine2021protein,alquraishi2021machine}, which has been trained to predict structures directly from sequence using an iterative approach \citep{alphafold2}. While the details are not fully available, it appears that a machine learning model has been optimized to predict directly the three-dimensional structure from sequence and trained using large sequence and structure databases.

We return now to the question of how such methods might be applied to IDPs, and note several obstacles that need to be overcome. First, the goal should not generally be to predict a single structure from the sequence, but rather an ensemble of dynamic structures. We here note in passing that some predictions provide multiple structures, but these generally represent the uncertainty of the prediction rather than the intrinsic heterogeneity and dynamics of the structure. Second, it is often difficult to generate high quality and deep MSAs of IDPs, in particular for those of low sequence complexity, or when conserved folded domains used to anchor alignments are lacking. Third, we do not have available a large number of structural ensembles that can be used to benchmark let alone train prediction methods. Thus, in contrast to the case for folded proteins where the PDB contains approximately 175.000 entries with protein structures, the PED contains ca. 200 ensembles.

Recent work has begun to develop approaches to understand the relationship between sequence and conformational ensembles of IDPs. In one such study, the concept of amino acid co-evolution was applied to predict contacts in IDPs from MSAs \citep{toth2016structured}. In several of the proteins analysed, the predicted contacts could be shown to coincide with key contacts observed within an IDP when bound in a complex to a folded protein \citep{toth2016structured}, thus demonstrating that the same principles that have been used so successfully for folded proteins have the potential to provide insight into IDPs, at least when they form complexes. In another study, we used similar sequence analyses of the disordered protein CsgA \citep{tian2015structure}. Here, we found a strong pattern of predicted contacts that corresponded to a folded amyloid-like state that CsgA forms. In this case, these contacts are preserved by evolution because CsgA forms a functional amyloid that is beneficial to the bacteria. While these and related studies suggest that sequence analysis might contain information that can be extracted to learn about the structures of IDPs, they have so far mostly revealed information about folded states that the IDPs might adopt or, perhaps, local secondary structure in the disordered states \citep{toth2016structured}. Here we note that there has been a substantial amount of work on predicting local structure in flexible peptides and proteins, but that our focus here is more generally on both local and global structures.

One of the limiting factors is also that we do not have a well-developed framework to discuss and quantify the relationship between sequence and ensemble properties. Indeed, as discussed above, protein disorder covers a continuum ranging from almost folded, but flexible and compact globules to chains that appear as statistical random coils. Because these proteins are best described in statistical terms, one approach is to bypass the three-dimensional structure all-together and predict key structural parameters directly from sequence. A number of such studies have focused on discovering the rules that govern the relationship between amino acid composition and patterning and, in particular, compaction of IDPs by parameterizing computational methods using experiments \citep{marsh2010sequence,hofmann2012polymer,das2013conformations,sawle2015theoretical,sorensen2019effective,zheng2020hydropathy,martin2020valence}. Recently, \cite{cohan2019information} developed a conceptual framework to examine such sequence-ensemble relationships more generally. Presumably, such approaches as well as more advanced computational methods and expanded sets of experimental data will be needed to predict structural properties beyond predicting compaction and local structure in IDPs.

\subsection{Beyond sequence alignments}
As discussed above, one of the main sources of information in current protein structure prediction comes from MSAs. While a detailed discussion of methods used to generate MSAs is beyond the scope of this paper, we note here that they are generally constructed based on the assumption of positional homology \citep{bawono2017multiple}, i.e. that a specific position in one sequence corresponds to a specific position in a homologous sequence. While this in turn is often the case for folded proteins, the situation in IDPs appears more complicated. Moreover, some IDPs diverge by gene duplication \citep{lee2008dif1} or are only found in some species \citep{rozen2015csnap}, and indels (insertion and deletions) are found to be frequent in IDPs \citep{light2013protein}, obscuring alignments further.

Recently, several ideas and methods from natural language processing have been applied to modelling, interpreting and predicting properties from protein sequences \citep{alley2019unified,rao2019evaluating,heinzinger2019modeling,ofer2021language}. While such methods are still difficult to interpret and expensive to train, they might help study proteins for which it is difficult to construct good alignments. Nevertheless, it still appears that MSAs contain substantial information that is not easily extracted from such language models \citep{rao2019evaluating}, and indeed combining the two can be advantageous \citep{rao2021msa}. 

Initial applications of language models to IDPs suggest that such models could be very useful in cases where one cannot construct accurate MSAs \citep{heinzinger2019modeling}. It will be interesting to explore whether these methods can be used to discover new rules that govern IDP sequences, and extract structural information from them. For example, as discussed above, properties such as the level of compaction and local structural motifs can to some extent already be predicted from sequence. It will, however, be interesting to explore how such methods can be improved---also for other properties such as long-range interactions or post-translational modifications---by developing new methods to represent sequences that are not based on the positional-conservation dogma that implicitly underlies many structure-prediction methods for folded proteins \citep{pritivsanac2019entropy,huihui2021intra}.

\subsection{Forward models for interpreting experimental data}
Many methods for predicting the structures of folded proteins are explicitly or implicitly based on the availability of thousands of labelled sequence-structure pairs in the PDB, used either for parameterization, training or validation. As discussed above, we have much fewer experimentally-derived conformational ensembles available for IDPs, and we here discuss how machine learning methods can provide potential advances in modelling IDPs.

While a plethora of methods exist for modelling conformational ensembles of IDPs, they are typically based on either biasing molecular simulations using experimental data or on selecting structures from a pre-generated ensemble to improve agreement with experiments \citep{mittag2007atomic,jensen2013describing,bonomi2017principles,orioli2020learn}. In these methods it is very important that the underlying dynamics and conformational averaging is treated correctly. For folded and rigid proteins one often transforms the experimental measurements into geometric restraints that are then applied during or after simulations. While this is possible for IDPs, a more general approach involves calculating experimental quantities from conformations and ensembles and comparing these to experiments. This calculation relies on so-called `forward models', i.e. algorithms to calculate experimentally-accessible quantities from conformational ensembles.

To give a concrete example, small-angle X-ray scattering (SAXS) experiments are often used to probe the compaction of an IDP, often quantified by the radius of gyration ($R_g$). One approach might therefore be to extract the $R_g$ from experimental SAXS data and generate a conformational ensemble with the same (average) $R_g$. Such an approach would, however ignore solvent contributions to the experimental measurements \citep{henriques2018calculation} as well as information from a wider range of scattering angles \citep{riback2017innovative,fuertes2017decoupling,zheng2018extended}. Also, when multiple sources of experimental data are used it is very important to treat errors and ensemble-averaging correctly \citep{ahmed2020computing}, and that becomes more difficult when working with quantities that are transformed values of the experimental measurements. Instead of using $R_g$, the more common approach is to use a forward model to calculate SAXS data from each conformation in an ensemble, and then compare the calculated average with experiments \citep{bernado2007structural}. A number of such forward models exist for SAXS experiments, that differ in how they treat solvent effects, as well as in accuracy and computational efficiency \citep{hub2018interpreting}. Importantly, different forward models may give different views of a conformational ensemble \citep{cordeiro2017disentangling,henriques2018calculation}, because---depending on the relationship between structure and measurement---different ensembles will be needed to agree with the experiments \citep{pesce2021refining}.

There are at least two different approaches to develop forward models, and these approaches can be combined. The first uses basic physical principles underlying the experiment to link structure and observable. Again using SAXS as an example, one of the most commonly used methods to calculate SAXS data from experiments (Crysol; \cite{svergun1995crysol}) calculates SAXS intensities from the scattering amplitude of the protein in vacuum as well as a model for the solvent contribution. The former is in turn based on empirically-derived form factors whereas the solvent contribution is parameterized using two parameters capturing the average solvent displaced by surface atoms and the excess density of the solvation layer. For folded and globular proteins, these two parameters are often fitted based either on a known structure or a model for the structure. Thus, the calculations of SAXS data from structural models may involve combining such a physical model while fitting one or a few empirical parameters against the experimental data.

Other forward models, such as for example methods that are used to calculate protein chemical shifts from protein structures are also often based on physical principles, but have a large number of parameters that need to be fit to experiments \citep{xu2001automated,shen2007protein,kohlhoff2009fast,han2011shiftx2}. This in turn is often based on data for folded proteins for which both high resolution structures and assigned chemical shifts are available. The mathematical function that connects structure and chemical shift is highly complex and has its roots in quantum mechanics. Thus, an alternative approach to express this relationship is to use neural networks \citep{meiler2003proshift,shen2010sparta+,li2020accurate,yang2020predicting}. One assumption underlying most of these approaches is that the experimental chemical shifts (which are time and ensemble averaged quantities) can be predicted accurately from a single structure. While that may be sufficient to study the rigid regions of folded proteins, other approaches may be needed to deal with more flexible parts \citep{lindorff2005interpreting,li2012ppm,christensen2013protein}. For IDPs where the chemical shifts represent a relatively broader ensemble of states and with more homogeneous chemical environments, further developments may be required to calculate accurate chemical shifts (relative to the small deviations from random coil values) and to extract structural information from these experiments \citep{crehuet2019bayesian}. 

Semi-empirical forward models such as those described above can be extremely difficult to develop for IDPs. This is because we rarely have sets of proteins for which we accurately know the conformational distribution derived independently from the set of measurements that one aims to develop a forward model for. Thus, most structures and ensembles determined for IDPs are implicitly based on forward models trained and validated on folded proteins. In the case of SAXS data this means, for example, that we often make the assumption that the solvation of a disordered protein is similar to that of a natively folded protein, and that the solvation properties are independent of the structure. While this may be true, this is very difficult to validate. One approach towards this goal may be to use more refined forward models to derive the ensembles \citep{hub2018interpreting,hermann2019saxs}, to reparameterize simplified models using such more refined methods \citep{henriques2018calculation}, or to refine ensembles and forward models in a self-consistent manner \citep{rieping2005inferential,brookes2016experimental,pesce2021refining}.

How can machine learning methods aid in the further development of forward models, and thus in our ability to derive conformational ensembles from experimental data? As described above, neural networks have already been used extensively to parameterize a function used to calculate chemical shifts, and we expect such methods will become refined and extended to a wider set of experiments. Machine learning methods have also been developed to extract shape information from SAXS experiments \citep{franke2018machine} though, to our knowledge, not as forward models. Similarly, a deep neural network based approach has been developed to process and extract structural information from electron paramagnetic resonance (EPR) experiments \citep{worswick2018deep}. Finally, a neural network was recently trained using quantum calculations to predict data from infrared  absorption spectroscopy \citep{ye2020machine}. Circular dichroism (CD) spectroscopy is widely used to study IDPs \citep{chemes2012circular}, yet calculating CD spectra from conformational ensembles of IDPs is difficult and generally based on `basic spectra' derived from folded proteins often via secondary structure classification \citep{nagy2019sesca}. We envisage that machine learning methods can aid in generalizing such approaches towards IDPs. In addition to the improved accuracy potentially afforded by such machine-learning-based forward models, they may also have other advantages such as rapid evaluation and differentiability, both of which can be important when determining conformational ensembles from experimental data.

\subsection{Improving energy functions for simulating IDPs}
Returning to the problem of predicting conformational properties and ensembles of IDPs from sequence we now explore how experiments and machine learning methods may be combined to improve conformational modelling. Ensembles generated either directly from molecular simulations or from integrative modelling using experiments are dependent on the quality of the physical models used in simulations \citep{orioli2020learn}. Thus improved force fields and energy functions both enable more accurate predictions of conformational properties from sequence, but also makes integrative methods more robust \citep{lindorff2009similarity, camilloni2012characterization,tiberti2015encore,larsen2020combining,ahmed2021refinement}. While molecular simulations may not be the most computationally efficient approach to predict conformational properties from sequence, it can serve as a benchmark and starting point for developing other approaches.

In recent years there have been substantial improvements in explicit solvent, all-atom force fields used to study the structure and dynamics of IDPs \citep{best2017computational,huang2018force,robustelli2018developing,mu2021recent}, and these improvements have been derived both by better quantum-level calculations and empirical fitting to experimental data. Conformational sampling of IDPs, in particular long IDPs or their complexes, remains a substantial challenge, and therefore implicit solvent or coarse-grained methods are sometimes used. These can in turn be parameterized using either bottom-up (based on more accurate models) or top-down (from experiments) approaches, or indeed a combination of the two \citep{noid2013perspective}. 

Some time ago we developed an automated approach to parameterize force fields based on experimental data and applied it to develop a coarse-grained model for IDPs \citep{norgaard2008experimental}. The basic idea, which had also been explored earlier for force field development \citep{njo1995determination,norrby1998automated,groth2001determination,bathe2003inverse}, is to sample force field parameter space and to optimize the parameters by comparing simulation results against experiments. Using a Bayesian framework it is possible to combine the experiments with other sources of information, and one may use reweighting techniques to speed up parameterization \citep{norgaard2008experimental}. In some sense, this approach can be considered a machine learning approach for learning force field parameters from experimental data. Later, similar ideas have been developed and applied to the problem of optimizing all-atom force fields against experimental data \citep{li2010nmr,piana2011robust,di2013automated,wang2014building,cesari2019fitting}. The ideas developed by \cite{norgaard2008experimental} have been extended and applied to larger sets of experimental data to construct coarse-grained \citep{chen2018learning,latham2019maximum,dannenhoffer2021data} and all-atom \citep{demerdash2019using} models for IDPs. In these approaches, the experimental data are used to refine or parameterize a fixed functional form for the force field and representation of the protein. Recently, a number of machine learning approaches have been developed and used both to construct force fields and to develop coarse-grained representations \citep{ruff2015camelot,husic2020coarse,gkeka2020machine,giulini2020information,yang2021construction}, and we expect such approaches could have a substantial impact on our ability to simulate IDPs at various resolutions.

The methods described above suggest that machine learning methods may be used both to improve our ability to calculate and interpret experimental observables and to parameterize computational models for IDPs directly against experiments. Common to both problems is the focus on interpreting and using the experimental measurements. This is key because the procedure when going from experimental measurement to conformational ensemble  involves approximations and loss of information. In the context of folded proteins, this is generally thought to be less of a concern, and the three dimensional coordinates are often a relatively good representation of the system and of the data. This in turn means that structure prediction methods can be trained or benchmarked on the protein structures (coordinates) rather than the experimental measurements used to derive them. We expect that this will not be the case for IDPs, and instead we suggest that machine learning methods for structure prediction should be benchmarked or trained directly on experimental data similarly to the force fields described above. Related, it is still an open question to what extent the complicated models used to predict protein structures from sequence internally represent the physics of proteins, and thus training models for structure prediction from experiments may end up being comparable to training molecular force fields.

\section{Towards predicting interactions and complexes}
\subsection{Identifying short linear motifs}
An noted above, the primary structures of disordered proteins are generally not very well conserved. Nevertheless, their sequences do carry important information about their function, clues to which can be derived from direct sequence analysis and alignments. Although complicated to perform, and often assisted by manual refinements and adjustments, it is still possible to construct multiple sequence alignments of disordered proteins and from these alignments identify conservation hotspots in otherwise poorly conserved regions. In such cases, few positions---as little as between two and five---are highly conserved across species and found to be distributed across a confined stretch of approximately a dozen residues. These conserved sequence stretches represent so-called Short Linear Motifs (SLiMs) \citep{neduva2005systematic,van2014short,jespersen2020emerging}. SLiMs are recurrent, and the same SLiM can be identified in different, seemingly unrelated proteins conferring binding to specific partner proteins or other biomolecules. They constitute interactions sites, and the conserved residues are essential contact points that form part of the complex interface, and are thus essential to IDPs and their interactome. Today, more than 2000 SLiMs have been identified and annotated, and more candidate SLiMs reported with many assembled in the Eukaryotic Linear Motif database (\url{elm.eu.org}; \cite{dinkel2012elm,kumar2020elm, gouw2020annotate}). It is, however, difficult to identify new SLiMs, define SLiM properties and specificity, and to annotate their functions. Below we discuss some areas where new experimental approaches and machine learning method may be integrated to shed further light on these problems.

One problem when applying machine learning methods to predict new instances from known SLiMs is that, typically, only a small number of experimentally verified cases are reported for each individual SLiM. This is mainly because methods for SLiM identification have been low-throughput and have relied mostly on bioinformatics approaches with subsequent biochemical and biological testing \citep{o2017structures}, or through integrating computation and medium throughput experiments \citep{zeke2015systematic}. More recently however, new high-throughput approaches have been used to define, expand and refine SLiMs. Examples include combining structure-based shape complementarity analysis and proteome-wide affinity purification mass spectrometry \citep{brauer2019leveraging} and proteomic peptide phage display (ProP-PD), a method for simultaneous proteome-scale identification of SLiM-mediated interactions and foot-printing of the binding region with amino acid resolution \citep{ivarsson2014large, sundell2018proteome}.  Recent work addressed $\approx$1,000,000 overlapping peptides covering the entire human disorderome in a single binding assay \citep{benz2021proteome}.

The generation of these large data sets provides new possibilities to train various types of prediction methods. Thus, a model has been trained to discriminate experimentally determined 14-3-3-binding SLiMs from non-binding phosphopeptides \citep{madeira201514} and a Random Forrest model was trained on a high-throughput phage display data set collected for low-specificity SLiM binding to S100A5 identifying recognition rules based on features of hydrophobicity and shape complementarity as primary determinants \citep{wheeler2020learning}. Likewise, prediction of binding regions in longer IDPs have been aided by the use of a trained bidirectional recurrent neural network, combining sequence, predicted secondary structures, Vina docking score and predicted disorder to improve the prediction \citep{khan2013predicting}. Thus, machine learning approaches may help identify features that define SLiM binding and specificity, and are often used together with 3D structures, as done e.g. for PDZ binding peptides \citep{kundu2014cluster}; a case where also more confident negatives could be included. Similar improvements in the number of reliable true-negatives were achieved in a reevaluation of a high-throughput binding data of SH2-pTyr interactions \citep{ronan2020new}. Currently such efforts are limited by a relative small number of large data sets, and further that larger scale experiments often address already known SLiMs. Another problem when developing prediction methods is the relatively low number of negative examples in many data sets, which has an impact on the number of false positives provided by the resulting models. Thus, ways to improve this issue are clearly needed. Once addressed, however, machine learning approaches could substantially further our understanding of SLiM-based interactions by enabling extraction of features of interaction that expand our view on sequence properties that determine SLiMs. Such features, which may also relate to conformational features, may help move beyond the expectation and limitation provided by a defined SLiM-sequence space. Indeed, SH2 domains, which are known to bind phospho-tyrosine ligands, have been shown to be able to also accommodate glutamates \citep{wallweber2014structural}, which would not be expected solely from the SLiM definition, and therefore not typically included in fragment based database designs for machine learning purposes \citep{plewczynski2005support}. Finally, results from machine learning approaches may have the further benefit of contributing to the development of new vocabulary to describe SLiM-based interactions and uncover novel rules for interactions by IDPs.

\subsection{Annotating function to short linear motifs}
Although many SLiMs have been classified, it has been estimated that the human proteome counts more than 100,000 SLiMs, leaving most SLiMs unidentified \citep{tompa2014million}. Needless to say, each newly discovered potential SLiM in a disordered protein needs experimental verification as well as annotation; a task that remains a huge effort and experimentally highly challenging. So, although identification of their presence may be relatively accessible, and even aided by machine learning approaches, functional annotation of SLiMs remains an obstacle. Current high-throughput approaches for functional annotation have used in vivo SLiM-dependent proximity labeling, and in silico modeling of motif determinants to uncover new interactors \citep{wigington2020systematic}, as well as ProP-PD \citep{ivarsson2014large, sundell2018proteome}.

There are, however, a number of complications that may make it difficult to apply machine learning methods to aid in annotating the function of newly discovered SLiMs. The same SLiM may in one protein be embedded in a sequence that folds to an $\alpha$-helix when bound, whereas in another protein, the same SLiM may form an extended structure or a $\beta$-strand when bound. One example is provided by a set of plant transcription factors that all bind to the $\alpha\alpha$-hub domain RST from RCD1 through the RST-binding SLiM. Here, the transcription factors individually form either a helix, an extended or disordered SLiM structures in the complexes \citep{o2017structures,bugge2018structure}. Thus, inherent to SLiMs is a certain plasticity in the position of the key conserved residues that form the critical contacts with the binding partner. Furthermore, the same sequence stretch within a disordered protein can have overlapping SLiMs and form biologically relevant complexes with different partners. There are several examples of this, for example the transcriptional activation domains of the tumor suppressor protein p53, which each have many different partners binding to the same overlapping region \citep{oldfield2008flexible,teilum2021new}.  

Once the target protein is known, additional complications can arise. One is SLiM `reversibility', in which two proteins with the same SLiM binds in opposite directions to the same partner, as shown for Sap25 and REST binding to Sin3-PAH1 \citep{swanson2004hbp1} and peptide binding to MHC class II molecules  \citep{gunther2010bidirectional}. This directly points to the SLiM context as carrying additional functional relevance \citep{stein2008contextual}. Indeed, it has been shown that the context may have both positive and negative effects on binding through charge attraction and repulsion \citep{palopoli2018short,prestel2019pcna}, and it may contribute to allosteric regulation \citep{garcia2010allostery,li2017genetically,o2017structures}.
Thus, the influence of context on SLiM-based interactions is emerging as functionally important and with a large potential relevant to drug targeting \citep{bugge2020interactions}. However, these flanking sequences and regions are often not conserved and are not resolved in experimental structures of the protein complexes---or even included in the experiments. Thus, these regions and their potential structural ensembles and conformational preferences cannot be extracted from the PDB and thus they currently constitute a data-gap for training purposes.

As the sequence properties of SLiMs are known only for a small fraction of the predicted SLiM-ome, there is a strong need for procedures that may enable the identification and annotation of SLiMs without extensive experimental efforts. Combined with the variability in the number of residues separating the key conserved sites within a SLiM, the possibility of being able to predict distance distributions of SLiM-based interactions, in which the possible contact points and special requirements could be mapped, would potentially be an important asset that may help facilitate functional annotation and even pinpoint relevant binding partners to address. While machine learning approaches seem like a promising approach, the elasticity of the SLiM sequence and the low conservation of the SLiM context would make a purely sequence-based approach difficult. Further information might be obtained from an MSA of the IDP and the binding partner \citep{skerker2008rewiring,burger2008accurate,weigt2009identification}, although the signal for contacts might be relatively weak and difficult to extract. Another problem that emerges is how to learn from sets of SLiMs that have been characterized in depth, and apply this knowledge to other sequences that have been probed much less.

One recently described approach to learn the rules for protein-peptide interactions is a bespoke machine-learning approach, termed hierarchical statistical mechanical modelling, which can be trained on families with abundant experimental data (structures and sequences) \citep{cunningham2020biophysical}. The approach learns a pseudo-energy function for  interactions relevant for binding, which can be transferred also to proteins for which less information is available. In this way, the approach provides an elegant example of how machine learning methods can be used to learn general rules of biophysics that enable transfer and predictions on a wider class of problems and systems.

Looking ahead, although many structures have been determined of complexes between folded domains and peptides representing SLiMs from disordered proteins, these structures have in most cases been solved in the absence of the flanking regions. As these regions can be highly relevant for binding specificity and affinity, it is important to develop approaches that take these sequences into account. At the moment, however, the functional and structural properties of flanking regions are poorly understood and rarely studied, making it difficult to develop prediction methods. Initially, it might be fruitful to compare the surface properties of the protein that binds the IDP (e.g. charge patterning and hydrophobicity) to the overall physico-chemical properties of the flanking regions. One approach towards such endeavours uses a sequence-based model of charge patterning to relate sequence to function \citep{huihui2021intra}. Eventually, and aided by the generation of data sets that includes longer peptides or full-length proteins, it may be possible to develop prediction methods that combine local and long-range interactions, perhaps using similar methods as when predicting effects of enhancers in gene regulation \citep{shlyueva2014transcriptional,avsec2021effective}.

\subsection{Complexes beyond SLiMs}
As most IDPs have large exposed surface areas with high conformational flexibility, they also have high potential for binding other proteins \citep{berlow2015functional,gao2018evolution}. IDP have thus shown remarkable structural and functional diversity in their complexes, ranging from complex formation through folding-upon-binding with interfaces of similar composition and properties as to those formed between folded complexes  \citep{rogers2014coupled,sugase2007mechanism,ievsmantavivcius2014helical,robustelli2020mechanism}, over complexes where the disordered partner remains dynamic to different extents \citep{brzovic2011acidic,tillu2021cavin1}. At the extreme end of the scale, highly dynamic complexes, which entirely lack the formation of stable secondary or tertiary structures, can form, for example between two highly and oppositely charged IDPs \citep{borgia2018extreme,schuler2020binding}. The dynamics retained in these complexes serve functional roles through very different mechanisms. Their dynamic properties lead to several mechanistic advantages such as complex partner exchange  \citep{berlow2017hypersensitive, berlow2019role}, facilitated dissociation via competitive substitution through formation of trimers \citep{sottini2020polyelectrolyte}, ensemble redistributions  \citep{henley2020unexpected}, and allosteric regulations \citep{milles2018ultraweak,hendus2019molecular}. Their malleability also confers other functional advantages to IDPs, one of which is the ability to bind multiple binding-partners as hubs, some at an overlapping site in competition, and some distributed along the chain leading to scaffolding and e.g. the formation of signalling complexes or transcriptional factories. How would machine learning methods aid in decomposing the role of disorder in functions of IDPs and what are the problems associated with this task? 

One of the first discoveries from studying disordered protein complexes were that they can fold upon binding, either to an already folded partner through one of two highly discussed mechanisms \citep{dogan2014binding,ievsmantavivcius2014helical,arai2015conformational} or through the occasional mutual folding of two disordered proteins \citep{demarest2002mutual,dogan2012fast}. Whereas folding-upon-binding of disordered regions at first may seem highly analogous to the process of protein folding, and hence in principle should be amiable to machine learning approaches to predict the structures of the complexes, there are however a number of obvious caveats to its direct use. First, even though the binding region may be known, it is not easy to predict from sequence alone, which part of the disordered protein will fold. Further, a continuum of disorder can exist both in the IDP alone and in a complex, and highly disordered complexes, by some termed fuzzy \citep{fuxreiter2012fuzzy,olsen2017behaviour}, may result in weak and near-stochastic interactions.

One such example is the activation domains of transcription factors \citep{erkine2018nonlinear}, whose properties were originally characterized as `acid blobs and negative noodles' \citep{sigler1988acid}. Recently, a number of multiplexed assays have been used to expand this view and study the functional requirements of the sequence properties of transcriptional activation domains \citep{staller2018high,ravarani2018high,erijman2020high,tycko2020high,sanborn2021simple,staller2021design}. These results confirm the original observations of a requirement for hydrophobic and negatively charged residues and provide additional information about the role of patterning. Further, the data can be used to train various sequence-based machine learning models for activity \citep{ravarani2018high,erijman2020high,sanborn2021simple,griffith2021parrot}. The results suggest that most functional variation can be explained solely by amino acid composition, but that there is additional signal from higher-order properties of the amino acid sequence \citep{erijman2020high}, thus highlighting the importance of generating sequence libraries with such properties in mind \citep{staller2018high}.

In addition to specific favourable interactions in a complex, binding by disordered proteins may also be driven by the use of entropy through other mechanisms \citep{pritivsanac2019entropy,flock2014controlling} such as via counter-ion release \citep{borgia2018extreme}, increased conformational flexibility in the complex or expansion of the surrounding disordered context \citep{heller2015targeting}. Prediction methods should ideally be able to quantify the remaining disorder after binding. Indeed, there are several examples of IDPs for which the bound state involves differently structured sub-populations of the complex, which all contribute to the specificity and selectivity in binding \citep{brzovic2011acidic,henley2020unexpected}, and there are complexes in which several contacts are made between the IDP and the folded partner, but where these dynamically and independently interchange (shuffle) just as in holding a hot potato \citep{perham1975self,hendus2016human}. While such examples provide difficult targets for prediction methods, they are also difficult to characterize by experiments, and thus there is limited data to train and benchmark on. 

\subsection{Biomolecular condensates}
Many IDPs have the ability to form multivalent interactions that are key for the ability to form so-called biomolecular condensates, either alone, with another IDP or in complex with folded domains or RNA. We refer the reader to recent reviews on the topic \citep{banani2017biomolecular,peran2020molecular,dignon2020biomolecular,choi2020physical}, and focus here mostly on the role of IDPs in forming such condensates and a set of problems where machine learning methods might help. 

Biomolecular condensates often form via the process of liquid-liquid phase separation (LLPS), and a central requirement for a molecule to form these structures is the ability to form multivalent interactions. In the context of IDPs, this can for example be a protein carrying multiple SLiMs that can bind to folded multidomain proteins \citep{li2012phase,bouchard2018cancer} or a set of amino acid residues within an IDP that can form sufficiently strong interactions between them \citep{wang2018molecular,martin2020valence}. Key areas for biophysical research include identifying the sequences and interactions that drive phase separation, identifying determinants of specificity in condensate formation, and elucidating the structural and dynamical features in biomolecular condensates. Before examining these questions, we stress that not all IDPs readily undergo LLPS, and that not all condensates involve IDPs.

Given the importance of IDP-IDP and SliM-target interactions, the methods discussed above for characterizing IDPs and SLiMs are also important for studying condensates. One key insight is that---due to the similarity between intramolecular interactions within IDPs and intermolecular interactions between IDPs---there is a correspondence between the propensity of an IDP to sample more compact structures and for it to undergo phase separation \citep{panagiotopoulos1998phase,lin2017phase,dignon2018relation,dignon2020biomolecular,choi2020physical,martin2020valence}. Thus, methods to predict compaction of IDPs or to parameterize simulation methods for isolated IDPs will also aid in studying phase separation of IDPs. Similarly, methods to predict SLiMs and their binding partners---and possibly the affinity of pairwise interactions---from sequence or from MSAs will aid in mapping the interactions that drive phase separation in these systems, and help to derive rules and features for their formation.

A number of databases have recently been created to collect information about proteins that undergo phase separation \citep{li2020llpsdb,li2020protein,meszaros2020phasepro,you2020phasepdb,ning2020drllps}. Such databases are now being used to develop prediction methods for phase separation \citep{vernon2018pi,hardenberg2020widespread,van2021predicting,raimondi2021silico,saar2021learning}, also with the aim of providing insight into the sequences and properties that are important for phase separation. In the same way as prediction methods for protein disorder have played a central role in understanding the role of disorder at the proteome level, such methods have the potential to do the same for biomolecular condensates \citep{vernon2018pi,hardenberg2020widespread}.

Moving ahead, it will be important to extend such databases and prediction methods with additional quantitative information on the propensity to phase separate, and to annotate more broadly what components or features are involved in the formation of condensates. In the same way as many proteins and peptides have been shown to form amyloid structures under some conditions, many proteins will likely undergo LLPS. Thus, in the same way as methods for predicting aggregation propensities have been trained on quantitative measurements of aggregation \citep{chiti2003rationalization,fernandez2004prediction,pawar2005prediction}, improvements in our ability to predict the propensity to undergo LLPS will likely involve fitting to or benchmarking against quantitative measurements of phase separation. Such analyses are already being performed with various coarse-grained simulation methods discussed above \citep{dignon2018sequence,martin2020valence,dignon2020biomolecular,choi2020physical,bremer2021deciphering}, but it may be difficult to scale these methods to proteomewide applications or to scan large numbers of components in heterotypic condensates. The relationship between intra- and inter-molecular interactions and the driving force for phase separation suggests that it might be possible to train sequence-based prediction models on single-chain properties and use these to predict the ability to undergo LLPS. Such methods have already provided a number of general rules about valency and patterning that appear promising for our ability to predict the propensity of proteins to undergo LLPS from their amino acid sequence \citep{martin2020valence,statt2020model,hazra2020charge,amin2020analytical,bremer2021deciphering}. Including the context, such as concentration, crowding and additional partners in heterotypic condensate formation in these models would be an important extension. The conformational landscape of IDPs is also dependent on a richness in protein post-translational modifications such as phosphorylations, methylations, sulfation, and lipidation, and e.g. phosphorylation and arginine methylation has been shown to affect the formation of condensates \citep{nott2015phase,monahan2017phosphorylation,lu2018phase,hofweber2018phase,hofweber2019friend}. Thus, predicting post-translational modifications and their effects on condensates would help provide additional insight into how condensates are regulated.

\section{Intrinsic disorder and human diseases}
Given their wide range of biological functions, it is not surprising that IDPs are involved in a number of human diseases \citep{uversky2009unfoldomics} including neurodegeneration \citep{uversky2015intrinsically} and in particular in cancer \citep{iakoucheva2002intrinsic,deiana2019intrinsically,meszaros2021mutations}. How may machine learning methods help understand the role of IDPs in disease?

While it appears a simple question to ask whether IDPs are enriched in a particular disease, answering this question requires accurate and unbiased predictions of protein disorder \citep{deiana2019intrinsically}. Thus, we need continuous development of databases and quantitative measures of protein disorder and assessment of prediction accuracy, as well as development of new prediction methods \citep{nielsen2019quality,dass2020odinpred,hatos2020disprot,necci2021critical}.

It is important to gain a better understanding of the molecular mechanisms underlying diseases involving IDPs. The expression of IDPs is tightly regulated, and misregulation may lead to disease \citep{babu2011intrinsically}. For folded proteins, it is well established that genetic missense variants may cause disease via a wide range of mechanisms including affecting both protein stability and interactions \citep{stefl2013molecular,sahni2015widespread,stein2019biophysical}. A substantial number of disease-causing variants are, however, located in regions of predicted disorder and are predicted to affect for example SLiMs \citep{vacic2012disease}. Thus, it is becoming clear that missense variants in IDPs can also lead to disease via perturbed interactions that either cause loss or gain of function  \citep{meyer2018mutations,li2019gain,wong2020protein}, including promoting the formation of fibrils and toxic oligomeric species.

Loss of protein stability arising from missense variants and resulting protein degradation is established to be a key mechanism underlying loss of function for many folded proteins \citep{casadio2011correlating,stein2019biophysical}, and indeed measurements or predictions of protein stability and abundance are useful for predicting loss of function \citep{matreyek2018multiplex,cagiada2021understanding}. While intrinsic thermodynamic stability of a folded state is not a meaningful quantity for IDPs, missense variants may still affect their cellular abundance. This may for example happen by mutations leading to impaired interactions and degradation, as exemplified by a missense variant in the IDR of the growth hormone receptor; a mutations leading to severe lung cancer \citep{chhabra2018growth}. Similarly, missense variants may lead to new interactions by SLiM appearance \citep{davey2015short,meyer2018mutations}, lack of degradation by interference with degrons, disorder-to-order formation \citep{vacic2012disease}, or changes in long-range interactions \citep{grazioli2019comparative}. In the latter example, machine learning techniques helped uncover differences in conformational dynamics from molecular dynamics trajectories of amyloid beta and the E22G disease variant, implicating their fibrillation into different morphologies. Thus, we need a better understanding both of the how IDPs are targeted for degradation and the sequence signals that determine cellular abundance \citep{van2014intrinsically}, and of how contact remodeling along the chain impacts the ensemble. Disordered regions may act as degradation signals (degrons) \citep{uversky2013most}, and new large-scale experiments are enabling a better understanding of the sequence and structural properties of degrons \citep{geffen2016mapping,koren2018eukaryotic}. We expect that such experiments will ultimately enable better predictions of the degradation and abundance of IDPs, and the effects of mutations on these properties.

One particularly important role of IDPs in disease may be in those that are involved in the formation of biomolecular condensates. A number of diseases have been associated with misregulation or formation of such condensates (as recently reviewed by others; \cite{aguzzi2016phase,shin2017liquid,elbaum2019matter,boija2021biomolecular,cai2021biomolecular,alberti2021biomolecular}), and thus a better understanding of the sequence properties that drive the formation of condensates will be important for predicting their role in disease \citep{tsang2020phase} as well as for targeting them pharmaceutically \citep{biesaga2021intrinsically}.

More generally, in order to better predict how variants in IDPs may cause disease, we need a clearer overview of the relationship between sequence, structural and dynamical properties, binding preferences and function. For folded proteins, analyses of conservation via MSAs are very powerful to predict whether a variant may cause disease \citep{riesselman2018deep,livesey2020using}, but as discussed above, constructing and analysing MSAs provide unique challenges for IDPs. Thus, we need new methods to leverage the increasingly growing sequence databases to predict the effects of sequence variation in IDPs \citep{zarin2019proteome,zhou2020idrmutpred,zarin2021identifying}, ultimately enabling targeting and drug development for combating diseases related to misregulation and dysfunction of disordered proteins. From an experimental point of view, multiplexed assays of variant effects (also sometimes called deep mutational scans) can provide key insights into both fundamental aspects of protein science \citep{fowler2014deep} and genotype-phenotype relationships and disease \citep{starita2017variant}. Such experiments are now also beginning to provide a more comprehensive view of the effects of amino acid subsitutions in IDPs such as the experiments on activation domains of transcription factors discussed above \citep{staller2018high,ravarani2018high,erijman2020high,tycko2020high,sanborn2021simple,staller2021design}, as well as experiments on a number of aggregation prone disordered proteins \citep{rogers2018nonproteinogenic, bolognesi2019mutational, gray2019elucidating, newberry2020deep,newberry2020robust,seuma2021genetic}.

\section{Outlook}
IDPs are an enormously broad class of molecules and together with IDRs they are involved in a wide range of biological functions. A key defining feature of IDPs and IDRs is something they do not have, namely a persistent three-dimensional structure. Thus, in many ways they are defined by being different from the globular and membrane proteins that in more than a century have been the central focus of much protein science. Indeed, when the first CASP experiment was performed in 1994 \citep{moult1995large} only few proteins where recognized as being intrinsically disordered and rarely was the conformational disorder linked to biological function.

In some ways, IDPs and IDRs are simpler than folded protein because their linear (primary) structure already provides much insight into their chemistry and ability to interact with other molecules. Thus, a number of computational methods have been developed to predict disorder from sequence and to identify local segments of the sequence that can bind to other molecules.

This apparent simplicity, however, can be deceiving. For folded proteins, the necessity to fold into a specific three-dimensional structure puts substantial restraints on the sequence and thus on evolution. Thus, MSAs of folded proteins often provide clues about specific residues and regions that are key to structure and function. Together with a large number of high resolution structures, this has lead to our ability to predict with increasing accuracy the structure of folded proteins. In contrast, while the sequences of many IDPs are conserved for function, this relationship is complex and different from that governing folded proteins, largely because their function is also coupled to their dynamics. It is interesting to speculate what protein science would have looked like if we had first discovered IDPs, and then later found sequences that fold into specific three-dimensional structures.

\begin{figure}[h]
\centering
\includegraphics[width=1.0\linewidth]{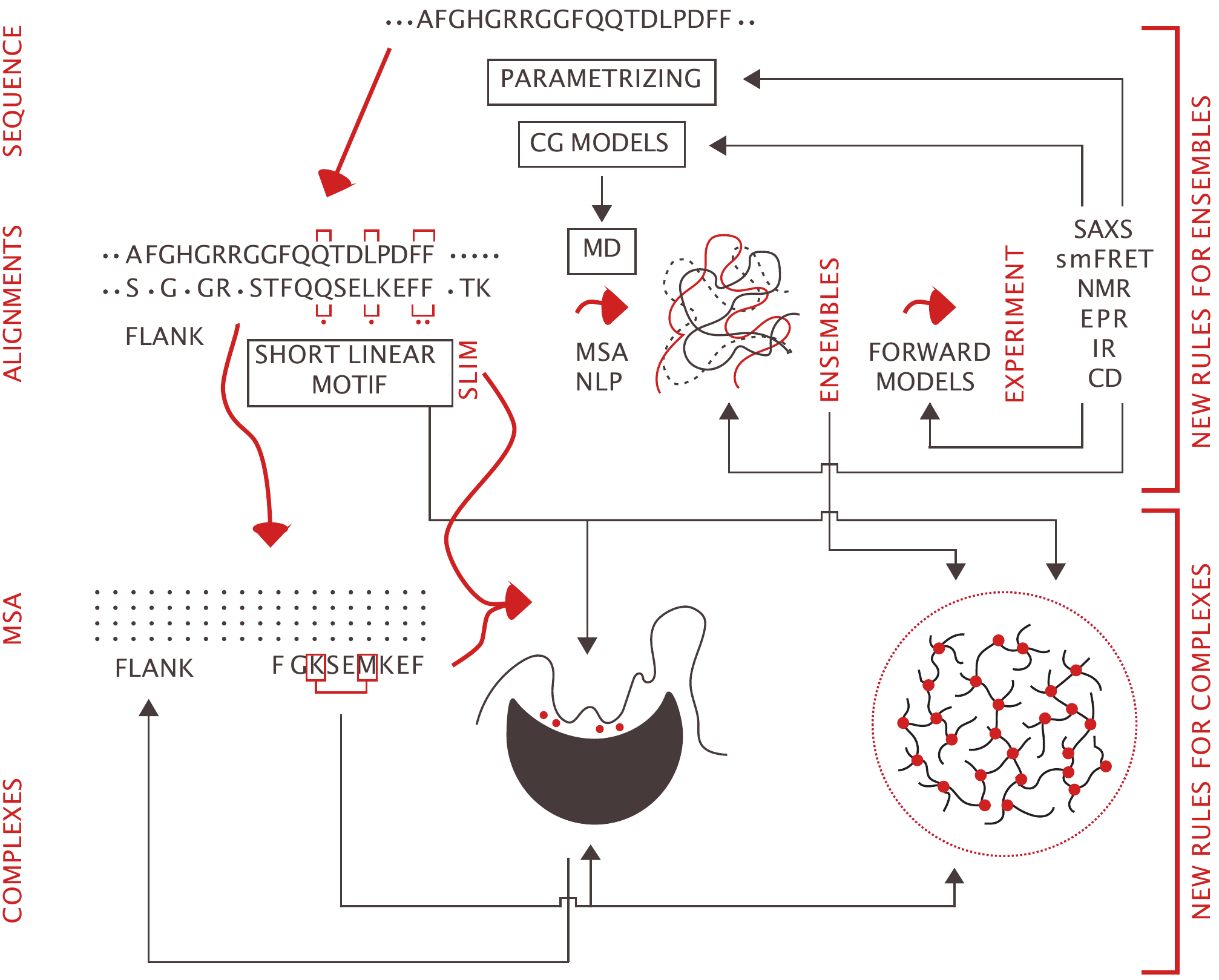}
\caption{Outlining connections between sequence, structure, dynamics and function of IDPs where implementing machine learning approaches could have a potential (indicated with black connectors). From sequences and sequence alignments, machine learning approaches may help extract conformational properties from poorly defined sequence alignments of IDPs. Machine learning may be used to improve methods for combining biophysical experiments (here illustrated by SAXS, NMR, smFRET, EPR, IR and CD) and computation for example by deriving better forward models, and helping parameterizing force fields for better coarse grained (CG) models of IDPs. Machine learning may also enable extraction of new SLiMs and annotation of their biological functions, and provide insight into and ability to predict how context and flanking region (flanks) contribute to IDP function. Machine learning may also help predict and understand properties important for the formation of biomolecular condensates, and how context plays roles in their formation and dissolution. Finally, but not illustrated here, the combination of these approaches may help assign pathogenicity to genetic variants of IDPs. United, machine learning in combination with bioinformatics, simulation, theory and experiments can  provide new rules for understanding IDP ensembles and IDP function. Jointly, such rules are necessary to enable the important decomposition of how mutations in IDPs may lead to disease states.}
\label{fig:fig2}
\end{figure}

Over the last 25 years, we have begun to understand the rules that govern the structural properties of IDPs, their interactions and their biological functions. Like for folded proteins, much insight has come from studying one system at a time, and computational methods are used to consolidate this into rules and predictions. In this review we have outlined a number of current problems in studies of IDPs, including our ability to characterize their structural preferences and interactions and our limitations in describing them. We have highlighted areas where machine learning and other computational methods have already had important impact, and new areas for further exploration (Fig.~\ref{fig:fig2}). Common to all is the tight interplay between experiment and computation. Particularly important is perhaps the realization that the two need to be developed together, with experiments being designed to inform computational methods, and computational algorithms developed, trained, and benchmarked using experiments. We look forward to see where these approaches will take the field. 

\section{Acknowledgments}
We acknowledge many discussions with our colleagues at the Structural Biology and NMR laboratory and Linderstr{\o}m-Lang Centre for Protein Science, and thank Tanja Mittag for comments on the manuscript. Asta B. Andersen is thanked for graphics support. Our research is supported by the Novo Nordisk Challenge Programmes REPIN (NNF18OC0033926; BBK) and PRISM (NNF18OC0033950; KLL), and the Lundbeck Foundation BRAINSTRUC initiative in structural biology (R155-2015-2666; KLL \& BBK).

\clearpage

\bibliography{references}

\begin{thebibliography}{264}
\providecommand{\natexlab}[1]{#1}
\providecommand{\urlprefix}{}
\providecommand{\doiprefix}{doi: }

\bibitem[{Aguzzi and Altmeyer(2016)Aguzzi, Adriano and Altmeyer,
  Matthias}]{aguzzi2016phase}
\textbf{\color{eLifeMediumGrey} Aguzzi A}, Altmeyer M.
\newblock Phase separation: linking cellular compartmentalization to disease.
\newblock Trends in cell biology.  2016; 26(7):547--558.

\bibitem[{Ahmed et~al.(2020)Ahmed, Mustapha Carab and Crehuet, Ramon and
  Lindorff-Larsen, Kresten}]{ahmed2020computing}
\textbf{\color{eLifeMediumGrey} Ahmed MC}, Crehuet R, Lindorff-Larsen K.
\newblock Computing, Analyzing, and Comparing the Radius of Gyration and
  Hydrodynamic Radius in Conformational Ensembles of Intrinsically Disordered
  Proteins.
\newblock In: \emph{Intrinsically Disordered Proteins} Springer; 2020.p.
  429--445.

\bibitem[{Ahmed et~al.(2021)Ahmed, Mustapha Carab and Skaanning, Line K and
  Jussupow, Alexander and Newcombe, Estella A and Kragelund, Birthe B and
  Camilloni, Carlo and Langkilde, Annette E and Lindorff-Larsen,
  Kresten}]{ahmed2021refinement}
\textbf{\color{eLifeMediumGrey} Ahmed MC}, Skaanning LK, Jussupow A, Newcombe
  EA, Kragelund BB, Camilloni C, Langkilde AE, Lindorff-Larsen K.
\newblock Refinement of $\alpha$-synuclein ensembles against SAXS data:
  Comparison of force fields and methods.
\newblock Frontiers in molecular biosciences.  2021; 8.

\bibitem[{Alberti and Hyman(2021)Alberti, Simon and Hyman, Anthony
  A}]{alberti2021biomolecular}
\textbf{\color{eLifeMediumGrey} Alberti S}, Hyman AA.
\newblock Biomolecular condensates at the nexus of cellular stress, protein
  aggregation disease and ageing.
\newblock Nature Reviews Molecular Cell Biology.  2021; p. 1--18.

\bibitem[{Alley et~al.(2019)Alley, Ethan C and Khimulya, Grigory and Biswas,
  Surojit and AlQuraishi, Mohammed and Church, George M}]{alley2019unified}
\textbf{\color{eLifeMediumGrey} Alley EC}, Khimulya G, Biswas S, AlQuraishi M,
  Church GM.
\newblock Unified rational protein engineering with sequence-based deep
  representation learning.
\newblock Nature methods.  2019; 16(12):1315--1322.

\bibitem[{AlQuraishi(2019)AlQuraishi, Mohammed}]{alquraishi2019end}
\textbf{\color{eLifeMediumGrey} AlQuraishi M}.
\newblock End-to-end differentiable learning of protein structure.
\newblock Cell systems.  2019; 8(4):292--301.

\bibitem[{AlQuraishi(2021)Mohammed AlQuraishi}]{alquraishi2021machine}
\textbf{\color{eLifeMediumGrey} AlQuraishi M}.
\newblock Machine learning in protein structure prediction.
\newblock Current Opinion in Chemical Biology.  2021; 65:1--8.

\bibitem[{Amin et~al.(2020)Amin, Alan N and Lin, Yi-Hsuan and Das, Suman and
  Chan, Hue Sun}]{amin2020analytical}
\textbf{\color{eLifeMediumGrey} Amin AN}, Lin YH, Das S, Chan HS.
\newblock Analytical theory for sequence-specific binary fuzzy complexes of
  charged intrinsically disordered proteins.
\newblock The Journal of Physical Chemistry B.  2020; 124(31):6709--6720.

\bibitem[{Anfinsen(1973)Anfinsen, Christian B}]{anfinsen1973principles}
\textbf{\color{eLifeMediumGrey} Anfinsen CB}.
\newblock Principles that govern the folding of protein chains.
\newblock Science.  1973; 181(4096):223--230.

\bibitem[{Arai et~al.(2015)Arai, Munehito and Sugase, Kenji and Dyson, H Jane
  and Wright, Peter E}]{arai2015conformational}
\textbf{\color{eLifeMediumGrey} Arai M}, Sugase K, Dyson HJ, Wright PE.
\newblock Conformational propensities of intrinsically disordered proteins
  influence the mechanism of binding and folding.
\newblock Proceedings of the National Academy of sciences.  2015;
  112(31):9614--9619.

\bibitem[{Avsec et~al.(2021)Avsec, Ziga and Agarwal, Vikram and Visentin,
  Daniel and Ledsam, Joseph R and Grabska-Barwinska, Agnieszka and Taylor, Kyle
  R and Assael, Yannis and Jumper, John and Kohli, Pushmeet and Kelley, David
  R}]{avsec2021effective}
\textbf{\color{eLifeMediumGrey} Avsec Z}, Agarwal V, Visentin D, Ledsam JR,
  Grabska-Barwinska A, Taylor KR, Assael Y, Jumper J, Kohli P, Kelley DR.
\newblock Effective gene expression prediction from sequence by integrating
  long-range interactions.
\newblock bioRxiv.  2021; .

\bibitem[{Babu et~al.(2011)Babu, M Madan and van der Lee, Robin and de Groot,
  Natalia Sanchez and Gsponer, J{\"o}rg}]{babu2011intrinsically}
\textbf{\color{eLifeMediumGrey} Babu MM}, van~der Lee R, de~Groot NS, Gsponer
  J.
\newblock Intrinsically disordered proteins: regulation and disease.
\newblock Current opinion in structural biology.  2011; 21(3):432--440.

\bibitem[{Balakrishnan et~al.(2011)Balakrishnan, Sivaraman and Kamisetty,
  Hetunandan and Carbonell, Jaime G and Lee, Su-In and Langmead, Christopher
  James}]{balakrishnan2011learning}
\textbf{\color{eLifeMediumGrey} Balakrishnan S}, Kamisetty H, Carbonell JG, Lee
  SI, Langmead CJ.
\newblock Learning generative models for protein fold families.
\newblock Proteins: Structure, Function, and Bioinformatics.  2011;
  79(4):1061--1078.

\bibitem[{Banani et~al.(2017)Banani, Salman F and Lee, Hyun O and Hyman,
  Anthony A and Rosen, Michael K}]{banani2017biomolecular}
\textbf{\color{eLifeMediumGrey} Banani SF}, Lee HO, Hyman AA, Rosen MK.
\newblock Biomolecular condensates: organizers of cellular biochemistry.
\newblock Nature reviews Molecular cell biology.  2017; 18(5):285--298.

\bibitem[{Bathe and Rutledge(2003)Bathe, Mark and Rutledge, Gregory
  C}]{bathe2003inverse}
\textbf{\color{eLifeMediumGrey} Bathe M}, Rutledge GC.
\newblock Inverse Monte Carlo procedure for conformation determination of
  macromolecules.
\newblock Journal of computational chemistry.  2003; 24(7):876--890.

\bibitem[{Bawono et~al.(2017)Bawono, Punto and Dijkstra, Maurits and Pirovano,
  Walter and Feenstra, Anton and Abeln, Sanne and Heringa,
  Jaap}]{bawono2017multiple}
\textbf{\color{eLifeMediumGrey} Bawono P}, Dijkstra M, Pirovano W, Feenstra A,
  Abeln S, Heringa J.
\newblock Multiple sequence alignment.
\newblock In: \emph{Bioinformatics} Springer; 2017.p. 167--189.

\bibitem[{Benz et~al.(2021)Benz, Caroline and Ali, Muhammad and Krystkowiak,
  Izabella and Simonetti, Leandro and Sayadi, Ahmed and Mihalic, Filip and
  Kliche, Johanna and Andersson, Eva and Jemth, Per and Davey, Norman E and
  others}]{benz2021proteome}
\textbf{\color{eLifeMediumGrey} Benz C}, Ali M, Krystkowiak I, Simonetti L,
  Sayadi A, Mihalic F, Kliche J, Andersson E, Jemth P, Davey NE, et~al.
\newblock Proteome-scale amino-acid resolution footprinting of protein-binding
  sites in the intrinsically disordered regions of the human proteome.
\newblock bioRxiv.  2021; .

\bibitem[{Berlow et~al.(2015)Berlow, Rebecca B and Dyson, H Jane and Wright,
  Peter E}]{berlow2015functional}
\textbf{\color{eLifeMediumGrey} Berlow RB}, Dyson HJ, Wright PE.
\newblock Functional advantages of dynamic protein disorder.
\newblock FEBS letters.  2015; 589(19):2433--2440.

\bibitem[{Berlow et~al.(2017)Berlow, Rebecca B and Dyson, H Jane and Wright,
  Peter E}]{berlow2017hypersensitive}
\textbf{\color{eLifeMediumGrey} Berlow RB}, Dyson HJ, Wright PE.
\newblock Hypersensitive termination of the hypoxic response by a disordered
  protein switch.
\newblock Nature.  2017; 543(7645):447--451.

\bibitem[{Berlow et~al.(2019)Berlow, Rebecca B and Martinez-Yamout, Maria A and
  Dyson, H Jane and Wright, Peter E}]{berlow2019role}
\textbf{\color{eLifeMediumGrey} Berlow RB}, Martinez-Yamout MA, Dyson HJ,
  Wright PE.
\newblock Role of backbone dynamics in modulating the interactions of
  disordered ligands with the TAZ1 domain of the CREB-binding protein.
\newblock Biochemistry.  2019; 58(10):1354--1362.

\bibitem[{Bernad{\'o} et~al.(2007)Bernad{\'o}, Pau and Mylonas, Efstratios and
  Petoukhov, Maxim V and Blackledge, Martin and Svergun, Dmitri
  I}]{bernado2007structural}
\textbf{\color{eLifeMediumGrey} Bernad{\'o} P}, Mylonas E, Petoukhov MV,
  Blackledge M, Svergun DI.
\newblock Structural characterization of flexible proteins using small-angle
  X-ray scattering.
\newblock Journal of the American Chemical Society.  2007; 129(17):5656--5664.

\bibitem[{Best(2017)Best, Robert B}]{best2017computational}
\textbf{\color{eLifeMediumGrey} Best RB}.
\newblock Computational and theoretical advances in studies of intrinsically
  disordered proteins.
\newblock Current opinion in structural biology.  2017; 42:147--154.

\bibitem[{Biesaga et~al.(2021)Biesaga, Mateusz and Frigol{\'e}-Vivas, Marta and
  Salvatella, Xavier}]{biesaga2021intrinsically}
\textbf{\color{eLifeMediumGrey} Biesaga M}, Frigol{\'e}-Vivas M, Salvatella X.
\newblock Intrinsically disordered proteins and biomolecular condensates as
  drug targets.
\newblock Current Opinion in Chemical Biology.  2021; 62:90--100.

\bibitem[{Boija et~al.(2021)Boija, Ann and Klein, Isaac A and Young, Richard
  A}]{boija2021biomolecular}
\textbf{\color{eLifeMediumGrey} Boija A}, Klein IA, Young RA.
\newblock Biomolecular condensates and cancer.
\newblock Cancer cell.  2021; .

\bibitem[{Bolognesi et~al.(2019)Bolognesi, Benedetta and Faure, Andre J and
  Seuma, Mireia and Schmiedel, J{\"o}rn M and Tartaglia, Gian Gaetano and
  Lehner, Ben}]{bolognesi2019mutational}
\textbf{\color{eLifeMediumGrey} Bolognesi B}, Faure AJ, Seuma M, Schmiedel JM,
  Tartaglia GG, Lehner B.
\newblock The mutational landscape of a prion-like domain.
\newblock Nature communications.  2019; 10(1):1--12.

\bibitem[{Bonomi et~al.(2017)Bonomi, Massimiliano and Heller, Gabriella T and
  Camilloni, Carlo and Vendruscolo, Michele}]{bonomi2017principles}
\textbf{\color{eLifeMediumGrey} Bonomi M}, Heller GT, Camilloni C, Vendruscolo
  M.
\newblock Principles of protein structural ensemble determination.
\newblock Current opinion in structural biology.  2017; 42:106--116.

\bibitem[{Borgia et~al.(2018)Borgia, Alessandro and Borgia, Madeleine B and
  Bugge, Katrine and Kissling, Vera M and Heidarsson, P{\'e}tur O and
  Fernandes, Catarina B and Sottini, Andrea and Soranno, Andrea and Buholzer,
  Karin J and Nettels, Daniel and Kragelund, Birthe B. and Best, Robert B. and
  Schuler, Benjamin}]{borgia2018extreme}
\textbf{\color{eLifeMediumGrey} Borgia A}, Borgia MB, Bugge K, Kissling VM,
  Heidarsson PO, Fernandes CB, Sottini A, Soranno A, Buholzer KJ, Nettels D,
  Kragelund BB, Best RB, Schuler B.
\newblock Extreme disorder in an ultrahigh-affinity protein complex.
\newblock Nature.  2018; 555(7694):61--66.

\bibitem[{Bouchard et~al.(2018)Bouchard, Jill J and Otero, Joel H and Scott,
  Daniel C and Szulc, Elzbieta and Martin, Erik W and Sabri, Nafiseh and
  Granata, Daniele and Marzahn, Melissa R and Lindorff-Larsen, Kresten and
  Salvatella, Xavier and others}]{bouchard2018cancer}
\textbf{\color{eLifeMediumGrey} Bouchard JJ}, Otero JH, Scott DC, Szulc E,
  Martin EW, Sabri N, Granata D, Marzahn MR, Lindorff-Larsen K, Salvatella X,
  et~al.
\newblock Cancer mutations of the tumor suppressor SPOP disrupt the formation
  of active, phase-separated compartments.
\newblock Molecular cell.  2018; 72(1):19--36.

\bibitem[{Brauer et~al.(2019)Brauer, Brooke L and Moon, Thomas M and Sheftic,
  Sarah R and Nasa, Isha and Page, Rebecca and Peti, Wolfgang and Kettenbach,
  Arminja N}]{brauer2019leveraging}
\textbf{\color{eLifeMediumGrey} Brauer BL}, Moon TM, Sheftic SR, Nasa I, Page
  R, Peti W, Kettenbach AN.
\newblock Leveraging new definitions of the LxVP SLiM to discover novel
  calcineurin regulators and substrates.
\newblock ACS chemical biology.  2019; 14(12):2672--2682.

\bibitem[{Bremer et~al.(2021)Bremer, Anne and Farag, Mina and Borcherds, Wade M
  and Peran, Ivan and Martin, Erik W and Pappu, Rohit V and Mittag,
  Tanja}]{bremer2021deciphering}
\textbf{\color{eLifeMediumGrey} Bremer A}, Farag M, Borcherds WM, Peran I,
  Martin EW, Pappu RV, Mittag T.
\newblock Deciphering how naturally occurring sequence features impact the
  phase behaviors of disordered prion-like domains.
\newblock bioRxiv.  2021; .

\bibitem[{Brookes and Head-Gordon(2016)Brookes, David H and Head-Gordon,
  Teresa}]{brookes2016experimental}
\textbf{\color{eLifeMediumGrey} Brookes DH}, Head-Gordon T.
\newblock Experimental inferential structure determination of ensembles for
  intrinsically disordered proteins.
\newblock Journal of the American Chemical Society.  2016; 138(13):4530--4538.

\bibitem[{Brzovic et~al.(2011)Brzovic, Peter S and Heikaus, Clemens C and
  Kisselev, Leonid and Vernon, Robert and Herbig, Eric and Pacheco, Derek and
  Warfield, Linda and Littlefield, Peter and Baker, David and Klevit, Rachel E
  and others}]{brzovic2011acidic}
\textbf{\color{eLifeMediumGrey} Brzovic PS}, Heikaus CC, Kisselev L, Vernon R,
  Herbig E, Pacheco D, Warfield L, Littlefield P, Baker D, Klevit RE, et~al.
\newblock The acidic transcription activator Gcn4 binds the mediator subunit
  Gal11/Med15 using a simple protein interface forming a fuzzy complex.
\newblock Molecular cell.  2011; 44(6):942--953.

\bibitem[{Bugge et~al.(2020)Bugge, Katrine and Brakti, Inna and Fernandes,
  Catarina B and Dreier, Jesper E and Lundsgaard, Jeppe E and Olsen, Johan G
  and Skriver, Karen and Kragelund, Birthe B}]{bugge2020interactions}
\textbf{\color{eLifeMediumGrey} Bugge K}, Brakti I, Fernandes CB, Dreier JE,
  Lundsgaard JE, Olsen JG, Skriver K, Kragelund BB.
\newblock Interactions by disorder--a matter of context.
\newblock Frontiers in Molecular Biosciences.  2020; 7.

\bibitem[{Bugge et~al.(2018)Bugge, Katrine and Staby, Lasse and Kemplen,
  Katherine R and O'Shea, Charlotte and Bendsen, Sidsel K and Jensen, Mikael K
  and Olsen, Johan G and Skriver, Karen and Kragelund, Birthe
  B}]{bugge2018structure}
\textbf{\color{eLifeMediumGrey} Bugge K}, Staby L, Kemplen KR, O'Shea C,
  Bendsen SK, Jensen MK, Olsen JG, Skriver K, Kragelund BB.
\newblock Structure of radical-induced cell death1 hub domain reveals a common
  $\alpha\alpha$-scaffold for disorder in transcriptional networks.
\newblock Structure.  2018; 26(5):734--746.

\bibitem[{Burger and Van~Nimwegen(2008)Burger, Lukas and Van Nimwegen,
  Erik}]{burger2008accurate}
\textbf{\color{eLifeMediumGrey} Burger L}, Van~Nimwegen E.
\newblock Accurate prediction of protein--protein interactions from sequence
  alignments using a Bayesian method.
\newblock Molecular systems biology.  2008; 4(1):165.

\bibitem[{Cagiada et~al.(2021)Cagiada, Matteo and Johansson, Kristoffer E and
  Valan{\v{c}}i{\=u}t{\.e}, Audron{\.e} and Nielsen, Sofie V and
  Hartmann-Petersen, Rasmus and Yang, Jun J and Fowler, Douglas M and Stein,
  Amelie and Lindorff-Larsen, Kresten}]{cagiada2021understanding}
\textbf{\color{eLifeMediumGrey} Cagiada M}, Johansson KE,
  Valan{\v{c}}i{\=u}t{\. e} A, Nielsen SV, Hartmann-Petersen R, Yang JJ, Fowler
  DM, Stein A, Lindorff-Larsen K.
\newblock Understanding the origins of loss of protein function by analyzing
  the effects of thousands of variants on activity and abundance.
\newblock Molecular Biology and Evolution.  2021; p. msab095.

\bibitem[{Cai et~al.(2021)Cai, Danfeng and Liu, Zhe and Lippincott-Schwartz,
  Jennifer}]{cai2021biomolecular}
\textbf{\color{eLifeMediumGrey} Cai D}, Liu Z, Lippincott-Schwartz J.
\newblock Biomolecular Condensates and Their Links to Cancer Progression.
\newblock Trends in biochemical sciences.  2021; .

\bibitem[{Camilloni et~al.(2012)Camilloni, Carlo and Robustelli, Paul and
  Simone, Alfonso De and Cavalli, Andrea and Vendruscolo,
  Michele}]{camilloni2012characterization}
\textbf{\color{eLifeMediumGrey} Camilloni C}, Robustelli P, Simone AD, Cavalli
  A, Vendruscolo M.
\newblock Characterization of the conformational equilibrium between the two
  major substates of RNase A using NMR chemical shifts.
\newblock Journal of the American Chemical Society.  2012; 134(9):3968--3971.

\bibitem[{Casadio et~al.(2011)Casadio, Rita and Vassura, Marco and Tiwari,
  Shalinee and Fariselli, Piero and Luigi Martelli,
  Pier}]{casadio2011correlating}
\textbf{\color{eLifeMediumGrey} Casadio R}, Vassura M, Tiwari S, Fariselli P,
  Luigi~Martelli P.
\newblock Correlating disease-related mutations to their effect on protein
  stability: A large-scale analysis of the human proteome.
\newblock Human mutation.  2011; 32(10):1161--1170.

\bibitem[{Cesari et~al.(2019)Cesari, Andrea and Bottaro, Sandro and
  Lindorff-Larsen, Kresten and Ban{\'a}\v{s}, Pavel and \v{S}poner,
  Ji\v{r}{\'{\i}} and Bussi, Giovanni}]{cesari2019fitting}
\textbf{\color{eLifeMediumGrey} Cesari A}, Bottaro S, Lindorff-Larsen K,
  Ban{\'a}\v{s} P, \v{S}poner J, Bussi G.
\newblock Fitting corrections to an RNA force field using experimental data.
\newblock Journal of chemical theory and computation.  2019; 15(6):3425--3431.

\bibitem[{Chemes et~al.(2012)Chemes, Luc{\'\i}a B and Alonso, Leonardo G and
  Noval, Mar{\'\i}a G and de Prat-Gay, Gonzalo}]{chemes2012circular}
\textbf{\color{eLifeMediumGrey} Chemes LB}, Alonso LG, Noval MG, de~Prat-Gay G.
\newblock Circular dichroism techniques for the analysis of intrinsically
  disordered proteins and domains.
\newblock In: \emph{Intrinsically disordered protein analysis} Springer;
  2012.p. 387--404.

\bibitem[{Chen et~al.(2018)Chen, Justin and Chen, Jiming and Pinamonti,
  Giovanni and Clementi, Cecilia}]{chen2018learning}
\textbf{\color{eLifeMediumGrey} Chen J}, Chen J, Pinamonti G, Clementi C.
\newblock Learning effective molecular models from experimental observables.
\newblock Journal of chemical theory and computation.  2018; 14(7):3849--3858.

\bibitem[{Chhabra et~al.(2018)Chhabra, Yash and Wong, Ho Yi and Nikolajsen, LF
  and Steinocher, Helena and Papadopulos, Andreas and Tunny, KA and Meunier, FA
  and Smith, AG and Kragelund, BB and Brooks, AJ and
  others}]{chhabra2018growth}
\textbf{\color{eLifeMediumGrey} Chhabra Y}, Wong HY, Nikolajsen L, Steinocher
  H, Papadopulos A, Tunny K, Meunier F, Smith A, Kragelund B, Brooks A, et~al.
\newblock A growth hormone receptor SNP promotes lung cancer by impairment of
  SOCS2-mediated degradation.
\newblock Oncogene.  2018; 37(4):489--501.

\bibitem[{Chiti et~al.(2003)Chiti, Fabrizio and Stefani, Massimo and Taddei,
  Niccolo and Ramponi, Giampietro and Dobson, Christopher
  M}]{chiti2003rationalization}
\textbf{\color{eLifeMediumGrey} Chiti F}, Stefani M, Taddei N, Ramponi G,
  Dobson CM.
\newblock Rationalization of the effects of mutations on peptide andprotein
  aggregation rates.
\newblock Nature.  2003; 424(6950):805--808.

\bibitem[{Choi et~al.(2020)Choi, Jeong-Mo and Holehouse, Alex S and Pappu,
  Rohit V}]{choi2020physical}
\textbf{\color{eLifeMediumGrey} Choi JM}, Holehouse AS, Pappu RV.
\newblock Physical principles underlying the complex biology of intracellular
  phase transitions.
\newblock Annual Review of Biophysics.  2020; 49:107--133.

\bibitem[{Christensen et~al.(2013)Christensen, Anders S and Linnet, Troels E
  and Borg, Mikael and Boomsma, Wouter and Lindorff-Larsen, Kresten and
  Hamelryck, Thomas and Jensen, Jan H}]{christensen2013protein}
\textbf{\color{eLifeMediumGrey} Christensen AS}, Linnet TE, Borg M, Boomsma W,
  Lindorff-Larsen K, Hamelryck T, Jensen JH.
\newblock Protein structure validation and refinement using amide proton
  chemical shifts derived from quantum mechanics.
\newblock PLoS One.  2013; 8(12):e84123.

\bibitem[{Cohan et~al.(2019)Cohan, Megan C and Ruff, Kiersten M and Pappu,
  Rohit V}]{cohan2019information}
\textbf{\color{eLifeMediumGrey} Cohan MC}, Ruff KM, Pappu RV.
\newblock Information theoretic measures for quantifying sequence--ensemble
  relationships of intrinsically disordered proteins.
\newblock Protein Engineering, Design and Selection.  2019; 32(4):191--202.

\bibitem[{wwPDB consortium(2019)}]{wwpdb2019protein}
\textbf{\color{eLifeMediumGrey} wwPDB consortium}.
\newblock Protein Data Bank: the single global archive for 3D macromolecular
  structure data.
\newblock Nucleic acids research.  2019; 47(D1):D520--D528.

\bibitem[{Cordeiro et~al.(2017)Cordeiro, Tiago N and Chen, Po-chia and De
  Biasio, Alfredo and Sibille, Nathalie and Blanco, Francisco J and Hub, Jochen
  S and Crehuet, Ramon and Bernad{\'o}, Pau}]{cordeiro2017disentangling}
\textbf{\color{eLifeMediumGrey} Cordeiro TN}, Chen Pc, De~Biasio A, Sibille N,
  Blanco FJ, Hub JS, Crehuet R, Bernad{\'o} P.
\newblock Disentangling polydispersity in the PCNA- p15PAF complex, a
  disordered, transient and multivalent macromolecular assembly.
\newblock Nucleic acids research.  2017; 45(3):1501--1515.

\bibitem[{Crehuet et~al.(2019)Crehuet, Ramon and Buigues, Pedro J and
  Salvatella, Xavier and Lindorff-Larsen, Kresten}]{crehuet2019bayesian}
\textbf{\color{eLifeMediumGrey} Crehuet R}, Buigues PJ, Salvatella X,
  Lindorff-Larsen K.
\newblock Bayesian-maximum-entropy reweighting of IDP ensembles based on NMR
  chemical shifts.
\newblock Entropy.  2019; 21(9):898.

\bibitem[{Cunningham et~al.(2020)Cunningham, Joseph M and Koytiger, Grigoriy
  and Sorger, Peter K and AlQuraishi, Mohammed}]{cunningham2020biophysical}
\textbf{\color{eLifeMediumGrey} Cunningham JM}, Koytiger G, Sorger PK,
  AlQuraishi M.
\newblock Biophysical prediction of protein--peptide interactions and signaling
  networks using machine learning.
\newblock Nature methods.  2020; 17(2):175--183.

\bibitem[{Dannenhoffer-Lafage and Best(2021)Dannenhoffer-Lafage, Thomas and
  Best, Robert B}]{dannenhoffer2021data}
\textbf{\color{eLifeMediumGrey} Dannenhoffer-Lafage T}, Best RB.
\newblock A Data-Driven Hydrophobicity Scale for Predicting Liquid--Liquid
  Phase Separation of Proteins.
\newblock The Journal of Physical Chemistry B.  2021; .

\bibitem[{Das and Pappu(2013)Das, Rahul K and Pappu, Rohit
  V}]{das2013conformations}
\textbf{\color{eLifeMediumGrey} Das RK}, Pappu RV.
\newblock Conformations of intrinsically disordered proteins are influenced by
  linear sequence distributions of oppositely charged residues.
\newblock Proceedings of the National Academy of Sciences.  2013;
  110(33):13392--13397.

\bibitem[{Dass et~al.(2020)Dass, Rupashree and Mulder, Frans AA and Nielsen,
  Jakob Toudahl}]{dass2020odinpred}
\textbf{\color{eLifeMediumGrey} Dass R}, Mulder FA, Nielsen JT.
\newblock ODiNPred: Comprehensive prediction of protein order and disorder.
\newblock Scientific reports.  2020; 10(1):1--16.

\bibitem[{Davey et~al.(2015)Davey, Norman E and Cyert, Martha S and Moses, Alan
  M}]{davey2015short}
\textbf{\color{eLifeMediumGrey} Davey NE}, Cyert MS, Moses AM.
\newblock Short linear motifs--ex nihilo evolution of protein regulation.
\newblock Cell Communication and Signaling.  2015; 13(1):1--15.

\bibitem[{Deiana et~al.(2019)Deiana, Antonio and Forcelloni, Sergio and
  Porrello, Alessandro and Giansanti, Andrea}]{deiana2019intrinsically}
\textbf{\color{eLifeMediumGrey} Deiana A}, Forcelloni S, Porrello A, Giansanti
  A.
\newblock Intrinsically disordered proteins and structured proteins with
  intrinsically disordered regions have different functional roles in the cell.
\newblock PloS one.  2019; 14(8):e0217889.

\bibitem[{Demarest et~al.(2002)Demarest, Stephen J and Martinez-Yamout, Maria
  and Chung, John and Chen, Hongwu and Xu, Wei and Dyson, H Jane and Evans,
  Ronald M and Wright, Peter E}]{demarest2002mutual}
\textbf{\color{eLifeMediumGrey} Demarest SJ}, Martinez-Yamout M, Chung J, Chen
  H, Xu W, Dyson HJ, Evans RM, Wright PE.
\newblock Mutual synergistic folding in recruitment of CBP/p300 by p160 nuclear
  receptor coactivators.
\newblock Nature.  2002; 415(6871):549--553.

\bibitem[{Demerdash et~al.(2019)Demerdash, Omar and Shrestha, Utsab R and
  Petridis, Loukas and Smith, Jeremy C and Mitchell, Julie C and Ramanathan,
  Arvind}]{demerdash2019using}
\textbf{\color{eLifeMediumGrey} Demerdash O}, Shrestha UR, Petridis L, Smith
  JC, Mitchell JC, Ramanathan A.
\newblock Using small-angle scattering data and parametric machine learning to
  optimize force field parameters for intrinsically disordered proteins.
\newblock Frontiers in molecular biosciences.  2019; 6:64.

\bibitem[{Di~Pierro and Elber(2013)Di Pierro, Michele and Elber,
  Ron}]{di2013automated}
\textbf{\color{eLifeMediumGrey} Di~Pierro M}, Elber R.
\newblock Automated optimization of potential parameters.
\newblock Journal of chemical theory and computation.  2013; 9(8):3311--3320.

\bibitem[{Dignon et~al.(2020)Dignon, Gregory L and Best, Robert B and Mittal,
  Jeetain}]{dignon2020biomolecular}
\textbf{\color{eLifeMediumGrey} Dignon GL}, Best RB, Mittal J.
\newblock Biomolecular phase separation: From molecular driving forces to
  macroscopic properties.
\newblock Annual review of physical chemistry.  2020; 71:53--75.

\bibitem[{Dignon et~al.(2018{\natexlab{a}})Dignon, Gregory L and Zheng, Wenwei
  and Best, Robert B and Kim, Young C and Mittal, Jeetain}]{dignon2018relation}
\textbf{\color{eLifeMediumGrey} Dignon GL}, Zheng W, Best RB, Kim YC, Mittal J.
\newblock Relation between single-molecule properties and phase behavior of
  intrinsically disordered proteins.
\newblock Proceedings of the National Academy of Sciences.  2018;
  115(40):9929--9934.

\bibitem[{Dignon et~al.(2018{\natexlab{b}})Dignon, Gregory L and Zheng, Wenwei
  and Kim, Young C and Best, Robert B and Mittal, Jeetain}]{dignon2018sequence}
\textbf{\color{eLifeMediumGrey} Dignon GL}, Zheng W, Kim YC, Best RB, Mittal J.
\newblock Sequence determinants of protein phase behavior from a coarse-grained
  model.
\newblock PLoS computational biology.  2018; 14(1):e1005941.

\bibitem[{Dinkel et~al.(2012)Dinkel, Holger and Michael, Sushama and
  Weatheritt, Robert J and Davey, Norman E and Van Roey, Kim and Altenberg,
  Brigitte and Toedt, Grischa and Uyar, Bora and Seiler, Markus and Budd, Aidan
  and others}]{dinkel2012elm}
\textbf{\color{eLifeMediumGrey} Dinkel H}, Michael S, Weatheritt RJ, Davey NE,
  Van~Roey K, Altenberg B, Toedt G, Uyar B, Seiler M, Budd A, et~al.
\newblock ELM—the database of eukaryotic linear motifs.
\newblock Nucleic acids research.  2012; 40(D1):D242--D251.

\bibitem[{Dogan et~al.(2014)Dogan, Jakob and Gianni, Stefano and Jemth,
  Per}]{dogan2014binding}
\textbf{\color{eLifeMediumGrey} Dogan J}, Gianni S, Jemth P.
\newblock The binding mechanisms of intrinsically disordered proteins.
\newblock Physical Chemistry Chemical Physics.  2014; 16(14):6323--6331.

\bibitem[{Dogan et~al.(2012)Dogan, Jakob and Schmidt, Tanja and Mu, Xin and
  Engstr{\"o}m, {\AA}ke and Jemth, Per}]{dogan2012fast}
\textbf{\color{eLifeMediumGrey} Dogan J}, Schmidt T, Mu X, Engstr{\"o}m {\AA},
  Jemth P.
\newblock Fast association and slow transitions in the interaction between two
  intrinsically disordered protein domains.
\newblock Journal of Biological Chemistry.  2012; 287(41):34316--34324.

\bibitem[{Eisenberg(2018)Eisenberg, David S}]{eisenberg2018hard}
\textbf{\color{eLifeMediumGrey} Eisenberg DS}.
\newblock How Hard It Is Seeing What Is in Front of Your Eyes.
\newblock Cell.  2018; 174(1):8--11.

\bibitem[{Elbaum-Garfinkle(2019)Elbaum-Garfinkle, Shana}]{elbaum2019matter}
\textbf{\color{eLifeMediumGrey} Elbaum-Garfinkle S}.
\newblock Matter over mind: Liquid phase separation and neurodegeneration.
\newblock Journal of Biological Chemistry.  2019; 294(18):7160--7168.

\bibitem[{Erijman et~al.(2020)Erijman, Ariel and Kozlowski, Lukasz and
  Sohrabi-Jahromi, Salma and Fishburn, James and Warfield, Linda and Schreiber,
  Jacob and Noble, William S and S{\"o}ding, Johannes and Hahn,
  Steven}]{erijman2020high}
\textbf{\color{eLifeMediumGrey} Erijman A}, Kozlowski L, Sohrabi-Jahromi S,
  Fishburn J, Warfield L, Schreiber J, Noble WS, S{\"o}ding J, Hahn S.
\newblock A High-Throughput screen for transcription activation domains reveals
  their sequence features and permits prediction by deep learning.
\newblock Molecular cell.  2020; 78(5):890--902.

\bibitem[{Erkine(2018)Erkine, Alexandre M}]{erkine2018nonlinear}
\textbf{\color{eLifeMediumGrey} Erkine AM}.
\newblock ‘Nonlinear’biochemistry of nucleosome detergents.
\newblock Trends in biochemical sciences.  2018; 43(12):951--959.

\bibitem[{Fernandez-Escamilla et~al.(2004)Fernandez-Escamilla, Ana-Maria and
  Rousseau, Frederic and Schymkowitz, Joost and Serrano,
  Luis}]{fernandez2004prediction}
\textbf{\color{eLifeMediumGrey} Fernandez-Escamilla AM}, Rousseau F,
  Schymkowitz J, Serrano L.
\newblock Prediction of sequence-dependent and mutational effects on the
  aggregation of peptides and proteins.
\newblock Nature biotechnology.  2004; 22(10):1302--1306.

\bibitem[{Flock et~al.(2014)Flock, Tilman and Weatheritt, Robert J and
  Latysheva, Natasha S and Babu, M Madan}]{flock2014controlling}
\textbf{\color{eLifeMediumGrey} Flock T}, Weatheritt RJ, Latysheva NS, Babu MM.
\newblock Controlling entropy to tune the functions of intrinsically disordered
  regions.
\newblock Current opinion in structural biology.  2014; 26:62--72.

\bibitem[{Fowler and Fields(2014)Fowler, Douglas M and Fields,
  Stanley}]{fowler2014deep}
\textbf{\color{eLifeMediumGrey} Fowler DM}, Fields S.
\newblock Deep mutational scanning: a new style of protein science.
\newblock Nature methods.  2014; 11(8):801--807.

\bibitem[{Franke et~al.(2018)Franke, Daniel and Jeffries, Cy M and Svergun,
  Dmitri I}]{franke2018machine}
\textbf{\color{eLifeMediumGrey} Franke D}, Jeffries CM, Svergun DI.
\newblock Machine learning methods for X-ray scattering data analysis from
  biomacromolecular solutions.
\newblock Biophysical journal.  2018; 114(11):2485--2492.

\bibitem[{Fuertes et~al.(2017)Fuertes, Gustavo and Banterle, Niccol{\`o} and
  Ruff, Kiersten M and Chowdhury, Aritra and Mercadante, Davide and Koehler,
  Christine and Kachala, Michael and Girona, Gemma Estrada and Milles, Sigrid
  and Mishra, Ankur and others}]{fuertes2017decoupling}
\textbf{\color{eLifeMediumGrey} Fuertes G}, Banterle N, Ruff KM, Chowdhury A,
  Mercadante D, Koehler C, Kachala M, Girona GE, Milles S, Mishra A, et~al.
\newblock Decoupling of size and shape fluctuations in heteropolymeric
  sequences reconciles discrepancies in SAXS vs. FRET measurements.
\newblock Proceedings of the National Academy of Sciences.  2017;
  114(31):E6342--E6351.

\bibitem[{Fuxreiter and Tompa(2012)Fuxreiter, Monika and Tompa,
  Peter}]{fuxreiter2012fuzzy}
\textbf{\color{eLifeMediumGrey} Fuxreiter M}, Tompa P.
\newblock Fuzzy complexes: a more stochastic view of protein function.
\newblock Fuzziness.  2012; p. 1--14.

\bibitem[{Gao et~al.(2018)Gao, Ang and Shrinivas, Krishna and Lepeudry, Paul
  and Suzuki, Hiroshi I and Sharp, Phillip A and Chakraborty, Arup
  K}]{gao2018evolution}
\textbf{\color{eLifeMediumGrey} Gao A}, Shrinivas K, Lepeudry P, Suzuki HI,
  Sharp PA, Chakraborty AK.
\newblock Evolution of weak cooperative interactions for biological
  specificity.
\newblock Proceedings of the National Academy of Sciences.  2018;
  115(47):E11053--E11060.

\bibitem[{Garcia-Pino et~al.(2010)Garcia-Pino, Abel and Balasubramanian,
  Sreeram and Wyns, Lode and Gazit, Ehud and De Greve, Henri and Magnuson, Roy
  D and Charlier, Daniel and van Nuland, Nico AJ and Loris,
  Remy}]{garcia2010allostery}
\textbf{\color{eLifeMediumGrey} Garcia-Pino A}, Balasubramanian S, Wyns L,
  Gazit E, De~Greve H, Magnuson RD, Charlier D, van Nuland NA, Loris R.
\newblock Allostery and intrinsic disorder mediate transcription regulation by
  conditional cooperativity.
\newblock Cell.  2010; 142(1):101--111.

\bibitem[{Geffen et~al.(2016)Geffen, Yifat and Appleboim, Alon and Gardner,
  Richard G and Friedman, Nir and Sadeh, Ronen and Ravid,
  Tommer}]{geffen2016mapping}
\textbf{\color{eLifeMediumGrey} Geffen Y}, Appleboim A, Gardner RG, Friedman N,
  Sadeh R, Ravid T.
\newblock Mapping the landscape of a eukaryotic degronome.
\newblock Molecular cell.  2016; 63(6):1055--1065.

\bibitem[{Giulini et~al.(2020)Giulini, Marco and Menichetti, Roberto and Shell,
  M Scott and Potestio, Raffaello}]{giulini2020information}
\textbf{\color{eLifeMediumGrey} Giulini M}, Menichetti R, Shell MS, Potestio R.
\newblock An Information-Theory-Based Approach for Optimal Model Reduction of
  Biomolecules.
\newblock Journal of chemical theory and computation.  2020; 16(11):6795--6813.

\bibitem[{Gkeka et~al.(2020)Gkeka, Paraskevi and Stoltz, Gabriel and Barati
  Farimani, Amir and Belkacemi, Zineb and Ceriotti, Michele and Chodera, John D
  and Dinner, Aaron R and Ferguson, Andrew L and Maillet, Jean-Bernard and
  Minoux, Herv{\'e} and others}]{gkeka2020machine}
\textbf{\color{eLifeMediumGrey} Gkeka P}, Stoltz G, Barati~Farimani A,
  Belkacemi Z, Ceriotti M, Chodera JD, Dinner AR, Ferguson AL, Maillet JB,
  Minoux H, et~al.
\newblock Machine learning force fields and coarse-grained variables in
  molecular dynamics: application to materials and biological systems.
\newblock Journal of Chemical Theory and Computation.  2020; 16(8):4757--4775.

\bibitem[{G{\"o}bel et~al.(1994)G{\"o}bel, Ulrike and Sander, Chris and
  Schneider, Reinhard and Valencia, Alfonso}]{gobel1994correlated}
\textbf{\color{eLifeMediumGrey} G{\"o}bel U}, Sander C, Schneider R, Valencia
  A.
\newblock Correlated mutations and residue contacts in proteins.
\newblock Proteins: Structure, Function, and Bioinformatics.  1994;
  18(4):309--317.

\bibitem[{Gouw et~al.(2020)Gouw, Marc and Alvarado-Valverde, Jes{\'u}s and
  {\v{C}}aly{\v{s}}eva, Jelena and Diella, Francesca and Kumar, Manjeet and
  Michael, Sushama and Van Roey, Kim and Dinkel, Holger and Gibson, Toby
  J}]{gouw2020annotate}
\textbf{\color{eLifeMediumGrey} Gouw M}, Alvarado-Valverde J,
  {\v{C}}aly{\v{s}}eva J, Diella F, Kumar M, Michael S, Van~Roey K, Dinkel H,
  Gibson TJ.
\newblock How to Annotate and Submit a Short Linear Motif to the Eukaryotic
  Linear Motif Resource.
\newblock In: \emph{Intrinsically Disordered Proteins} Springer; 2020.p.
  73--102.

\bibitem[{Gray et~al.(2019)Gray, Vanessa E and Sitko, Katherine and Kameni,
  Floriane Z Ngako and Williamson, Miriam and Stephany, Jason J and Hasle,
  Nicholas and Fowler, Douglas M}]{gray2019elucidating}
\textbf{\color{eLifeMediumGrey} Gray VE}, Sitko K, Kameni FZN, Williamson M,
  Stephany JJ, Hasle N, Fowler DM.
\newblock Elucidating the molecular determinants of A$\beta$ aggregation with
  deep mutational scanning.
\newblock G3: Genes, Genomes, Genetics.  2019; 9(11):3683--3689.

\bibitem[{Grazioli et~al.(2019)Grazioli, Gianmarc and Martin, Rachel W and
  Butts, Carter T}]{grazioli2019comparative}
\textbf{\color{eLifeMediumGrey} Grazioli G}, Martin RW, Butts CT.
\newblock Comparative exploratory analysis of intrinsically disordered protein
  dynamics using machine learning and network analytic methods.
\newblock Frontiers in molecular biosciences.  2019; 6:42.

\bibitem[{Griffith and Holehouse(2021)Griffith, Daniel and Holehouse, Alex
  S}]{griffith2021parrot}
\textbf{\color{eLifeMediumGrey} Griffith D}, Holehouse AS.
\newblock PARROT: a flexible recurrent neural network framework for analysis of
  large protein datasets.
\newblock bioRxiv.  2021; .

\bibitem[{Groth et~al.(2001)Groth, Ma{\l}gorzata and Malicka, Joanna and
  Rodziewicz-Motowid{\l}o, Sylwia and Czaplewski, Cezary and Klaudel, Lidia and
  Wiczk, Wies{\l}aw and Liwo, Adam}]{groth2001determination}
\textbf{\color{eLifeMediumGrey} Groth M}, Malicka J, Rodziewicz-Motowid{\l}o S,
  Czaplewski C, Klaudel L, Wiczk W, Liwo A.
\newblock Determination of conformational equilibrium of peptides in solution
  by NMR spectroscopy and theoretical conformational analysis: Application to
  the calibration of mean-field solvation models.
\newblock Peptide Science: Original Research on Biomolecules.  2001;
  60(2):79--95.

\bibitem[{G{\"u}nther et~al.(2010)G{\"u}nther, Sebastian and Schlundt, Andreas
  and Sticht, Jana and Roske, Yvette and Heinemann, Udo and Wiesm{\"u}ller,
  Karl-Heinz and Jung, G{\"u}nther and Falk, Kirsten and R{\"o}tzschke, Olaf
  and Freund, Christian}]{gunther2010bidirectional}
\textbf{\color{eLifeMediumGrey} G{\"u}nther S}, Schlundt A, Sticht J, Roske Y,
  Heinemann U, Wiesm{\"u}ller KH, Jung G, Falk K, R{\"o}tzschke O, Freund C.
\newblock Bidirectional binding of invariant chain peptides to an MHC class II
  molecule.
\newblock Proceedings of the National Academy of Sciences.  2010;
  107(51):22219--22224.

\bibitem[{Han et~al.(2011)Han, Beomsoo and Liu, Yifeng and Ginzinger, Simon W
  and Wishart, David S}]{han2011shiftx2}
\textbf{\color{eLifeMediumGrey} Han B}, Liu Y, Ginzinger SW, Wishart DS.
\newblock SHIFTX2: significantly improved protein chemical shift prediction.
\newblock Journal of biomolecular NMR.  2011; 50(1):43.

\bibitem[{Hardenberg et~al.(2020)Hardenberg, Maarten and Horvath, Attila and
  Ambrus, Viktor and Fuxreiter, Monika and Vendruscolo,
  Michele}]{hardenberg2020widespread}
\textbf{\color{eLifeMediumGrey} Hardenberg M}, Horvath A, Ambrus V, Fuxreiter
  M, Vendruscolo M.
\newblock Widespread occurrence of the droplet state of proteins in the human
  proteome.
\newblock Proceedings of the National Academy of Sciences.  2020;
  117(52):33254--33262.

\bibitem[{Hatos et~al.(2020)Hatos, Andr{\'a}s and Hajdu-Solt{\'e}sz,
  Borb{\'a}la and Monzon, Alexander M and Palopoli, Nicolas and {\'A}lvarez,
  Luc{\'\i}a and Aykac-Fas, Burcu and Bassot, Claudio and Ben{\'\i}tez,
  Guillermo I and Bevilacqua, Martina and Chasapi, Anastasia and
  others}]{hatos2020disprot}
\textbf{\color{eLifeMediumGrey} Hatos A}, Hajdu-Solt{\'e}sz B, Monzon AM,
  Palopoli N, {\'A}lvarez L, Aykac-Fas B, Bassot C, Ben{\'\i}tez GI, Bevilacqua
  M, Chasapi A, et~al.
\newblock DisProt: intrinsic protein disorder annotation in 2020.
\newblock Nucleic acids research.  2020; 48(D1):D269--D276.

\bibitem[{Hazra and Levy(2020)Hazra, Milan Kumar and Levy,
  Yaakov}]{hazra2020charge}
\textbf{\color{eLifeMediumGrey} Hazra MK}, Levy Y.
\newblock Charge pattern affects the structure and dynamics of polyampholyte
  condensates.
\newblock Physical Chemistry Chemical Physics.  2020; 22(34):19368--19375.

\bibitem[{Heinzinger et~al.(2019)Heinzinger, Michael and Elnaggar, Ahmed and
  Wang, Yu and Dallago, Christian and Nechaev, Dmitrii and Matthes, Florian and
  Rost, Burkhard}]{heinzinger2019modeling}
\textbf{\color{eLifeMediumGrey} Heinzinger M}, Elnaggar A, Wang Y, Dallago C,
  Nechaev D, Matthes F, Rost B.
\newblock Modeling aspects of the language of life through transfer-learning
  protein sequences.
\newblock BMC bioinformatics.  2019; 20(1):1--17.

\bibitem[{Heller et~al.(2015)Heller, Gabriella T and Sormanni, Pietro and
  Vendruscolo, Michele}]{heller2015targeting}
\textbf{\color{eLifeMediumGrey} Heller GT}, Sormanni P, Vendruscolo M.
\newblock Targeting disordered proteins with small molecules using entropy.
\newblock Trends in biochemical sciences.  2015; 40(9):491--496.

\bibitem[{Hendus-Altenburger et~al.(2016)Hendus-Altenburger, Ruth and
  Pedraz-Cuesta, Elena and Olesen, Christina W and Papaleo, Elena and Schnell,
  Jeff A and Hopper, Jonathan TS and Robinson, Carol V and Pedersen, Stine F
  and Kragelund, Birthe B}]{hendus2016human}
\textbf{\color{eLifeMediumGrey} Hendus-Altenburger R}, Pedraz-Cuesta E, Olesen
  CW, Papaleo E, Schnell JA, Hopper JT, Robinson CV, Pedersen SF, Kragelund BB.
\newblock The human Na+/H+ exchanger 1 is a membrane scaffold protein for
  extracellular signal-regulated kinase 2.
\newblock BMC biology.  2016; 14(1):1--17.

\bibitem[{Hendus-Altenburger et~al.(2019)Hendus-Altenburger, Ruth and Wang,
  Xinru and Sj{\o}gaard-Frich, Lise M and Pedraz-Cuesta, Elena and Sheftic,
  Sarah R and Bends{\o}e, Anne H and Page, Rebecca and Kragelund, Birthe B and
  Pedersen, Stine F and Peti, Wolfgang}]{hendus2019molecular}
\textbf{\color{eLifeMediumGrey} Hendus-Altenburger R}, Wang X,
  Sj{\o}gaard-Frich LM, Pedraz-Cuesta E, Sheftic SR, Bends{\o}e AH, Page R,
  Kragelund BB, Pedersen SF, Peti W.
\newblock Molecular basis for the binding and selective dephosphorylation of
  Na+/H+ exchanger 1 by calcineurin.
\newblock Nature communications.  2019; 10(1):1--13.

\bibitem[{Henley et~al.(2020)Henley, Matthew J and Linhares, Brian M and
  Morgan, Brittany S and Cierpicki, Tomasz and Fierke, Carol A and Mapp, Anna
  K}]{henley2020unexpected}
\textbf{\color{eLifeMediumGrey} Henley MJ}, Linhares BM, Morgan BS, Cierpicki
  T, Fierke CA, Mapp AK.
\newblock Unexpected specificity within dynamic transcriptional
  protein--protein complexes.
\newblock Proceedings of the National Academy of Sciences.  2020;
  117(44):27346--27353.

\bibitem[{Henriques et~al.(2018)Henriques, Jo{\~a}o and Arleth, Lise and
  Lindorff-Larsen, Kresten and Skep{\"o}, Marie}]{henriques2018calculation}
\textbf{\color{eLifeMediumGrey} Henriques J}, Arleth L, Lindorff-Larsen K,
  Skep{\"o} M.
\newblock On the calculation of SAXS profiles of folded and intrinsically
  disordered proteins from computer simulations.
\newblock Journal of molecular biology.  2018; 430(16):2521--2539.

\bibitem[{Hermann and Hub(2019)Hermann, Markus R and Hub, Jochen
  S}]{hermann2019saxs}
\textbf{\color{eLifeMediumGrey} Hermann MR}, Hub JS.
\newblock SAXS-restrained ensemble simulations of intrinsically disordered
  proteins with commitment to the principle of maximum entropy.
\newblock Journal of chemical theory and computation.  2019; 15(9):5103--5115.

\bibitem[{Hofmann et~al.(2012)Hofmann, Hagen and Soranno, Andrea and Borgia,
  Alessandro and Gast, Klaus and Nettels, Daniel and Schuler,
  Benjamin}]{hofmann2012polymer}
\textbf{\color{eLifeMediumGrey} Hofmann H}, Soranno A, Borgia A, Gast K,
  Nettels D, Schuler B.
\newblock Polymer scaling laws of unfolded and intrinsically disordered
  proteins quantified with single-molecule spectroscopy.
\newblock Proceedings of the National Academy of Sciences.  2012;
  109(40):16155--16160.

\bibitem[{Hofweber and Dormann(2019)Hofweber, Mario and Dormann,
  Dorothee}]{hofweber2019friend}
\textbf{\color{eLifeMediumGrey} Hofweber M}, Dormann D.
\newblock Friend or foe—Post-translational modifications as regulators of
  phase separation and RNP granule dynamics.
\newblock Journal of Biological Chemistry.  2019; 294(18):7137--7150.

\bibitem[{Hofweber et~al.(2018)Hofweber, Mario and Hutten, Saskia and
  Bourgeois, Benjamin and Spreitzer, Emil and Niedner-Boblenz, Annika and
  Schifferer, Martina and Ruepp, Marc-David and Simons, Mikael and Niessing,
  Dierk and Madl, Tobias and others}]{hofweber2018phase}
\textbf{\color{eLifeMediumGrey} Hofweber M}, Hutten S, Bourgeois B, Spreitzer
  E, Niedner-Boblenz A, Schifferer M, Ruepp MD, Simons M, Niessing D, Madl T,
  et~al.
\newblock Phase separation of FUS is suppressed by its nuclear import receptor
  and arginine methylation.
\newblock Cell.  2018; 173(3):706--719.

\bibitem[{Huang and MacKerell~Jr(2018)Huang, Jing and MacKerell Jr, Alexander
  D}]{huang2018force}
\textbf{\color{eLifeMediumGrey} Huang J}, MacKerell~Jr AD.
\newblock Force field development and simulations of intrinsically disordered
  proteins.
\newblock Current opinion in structural biology.  2018; 48:40--48.

\bibitem[{Hub(2018)Hub, Jochen S}]{hub2018interpreting}
\textbf{\color{eLifeMediumGrey} Hub JS}.
\newblock Interpreting solution X-ray scattering data using molecular
  simulations.
\newblock Current opinion in structural biology.  2018; 49:18--26.

\bibitem[{Huihui and Ghosh(2021)Huihui, J and Ghosh, K}]{huihui2021intra}
\textbf{\color{eLifeMediumGrey} Huihui J}, Ghosh K.
\newblock Intra-chain interaction topology can identify functionally similar
  Intrinsically Disordered Proteins.
\newblock Biophysical Journal.  2021; p. 1860--1868.

\bibitem[{Husic et~al.(2020)Husic, Brooke E and Charron, Nicholas E and Lemm,
  Dominik and Wang, Jiang and P{\'e}rez, Adri{\`a} and Majewski, Maciej and
  Kr{\"a}mer, Andreas and Chen, Yaoyi and Olsson, Simon and de Fabritiis,
  Gianni and others}]{husic2020coarse}
\textbf{\color{eLifeMediumGrey} Husic BE}, Charron NE, Lemm D, Wang J,
  P{\'e}rez A, Majewski M, Kr{\"a}mer A, Chen Y, Olsson S, de~Fabritiis G,
  et~al.
\newblock Coarse graining molecular dynamics with graph neural networks.
\newblock The Journal of Chemical Physics.  2020; 153(19):194101.

\bibitem[{Iakoucheva et~al.(2002)Iakoucheva, Lilia M and Brown, Celeste J and
  Lawson, J David and Obradovi{\'c}, Zoran and Dunker, A
  Keith}]{iakoucheva2002intrinsic}
\textbf{\color{eLifeMediumGrey} Iakoucheva LM}, Brown CJ, Lawson JD,
  Obradovi{\'c} Z, Dunker AK.
\newblock Intrinsic disorder in cell-signaling and cancer-associated proteins.
\newblock Journal of molecular biology.  2002; 323(3):573--584.

\bibitem[{Ie{\v{s}}mantavi{\v{c}}ius et~al.(2014)Ie{\v{s}}mantavi{\v{c}}ius,
  Vytautas and Dogan, Jakob and Jemth, Per and Teilum, Kaare and Kjaergaard,
  Magnus}]{ievsmantavivcius2014helical}
\textbf{\color{eLifeMediumGrey} Ie{\v{s}}mantavi{\v{c}}ius V}, Dogan J, Jemth
  P, Teilum K, Kjaergaard M.
\newblock Helical propensity in an intrinsically disordered protein accelerates
  ligand binding.
\newblock Angewandte Chemie International Edition.  2014; 53(6):1548--1551.

\bibitem[{Ivarsson et~al.(2014)Ivarsson, Ylva and Arnold, Roland and
  McLaughlin, Megan and Nim, Satra and Joshi, Rakesh and Ray, Debashish and
  Liu, Bernard and Teyra, Joan and Pawson, Tony and Moffat, Jason and
  others}]{ivarsson2014large}
\textbf{\color{eLifeMediumGrey} Ivarsson Y}, Arnold R, McLaughlin M, Nim S,
  Joshi R, Ray D, Liu B, Teyra J, Pawson T, Moffat J, et~al.
\newblock Large-scale interaction profiling of PDZ domains through proteomic
  peptide-phage display using human and viral phage peptidomes.
\newblock Proceedings of the National Academy of Sciences.  2014;
  111(7):2542--2547.

\bibitem[{Jensen et~al.(2013)Jensen, Malene Ringkj{\o}bing and Ruigrok, Rob WH
  and Blackledge, Martin}]{jensen2013describing}
\textbf{\color{eLifeMediumGrey} Jensen MR}, Ruigrok RW, Blackledge M.
\newblock Describing intrinsically disordered proteins at atomic resolution by
  NMR.
\newblock Current opinion in structural biology.  2013; 23(3):426--435.

\bibitem[{Jespersen and Barbar(2020)Jespersen, Nathan and Barbar,
  Elisar}]{jespersen2020emerging}
\textbf{\color{eLifeMediumGrey} Jespersen N}, Barbar E.
\newblock Emerging features of linear motif-binding Hub proteins.
\newblock Trends in biochemical sciences.  2020; 45(5):375--384.

\bibitem[{Jumper et~al.(2020)Jumper, John and Evans, Richard and Pritzel,
  Alexander and Green, Tim and Figurnov, Michael and Tunyasuvunakool, Kathryn
  and Ronneberger, Olaf and Bates, Russ and \v{Z}\'{\i}dek, Augustin and
  Bridgland, Alex and Meyer, Clemens and Kohl, Simon A A and Potapenko, Anna
  and Ballard, Andrew J and Cowie, Andrew and Romera-Paredes, Bernardino and
  Nikolov, Stanislav and Jain, Rishub and Adler, Jonas and Back, Trevor and
  Petersen, Stig and Reiman, David and Steinegger, Martin and Pacholska,
  Michalina and Silver, David and Vinyals, Oriol and Senior, Andrew W and
  Kavukcuoglu, Koray and Kohli, Pushmeet and Hassabis, Demis}]{alphafold2}
\textbf{\color{eLifeMediumGrey} Jumper J}, Evans R, Pritzel A, Green T,
  Figurnov M, Tunyasuvunakool K, Ronneberger O, Bates R, \v{Z}\'{\i}dek A,
  Bridgland A, Meyer C, Kohl SAA, Potapenko A, Ballard AJ, Cowie A,
  Romera-Paredes B, Nikolov S, Jain R, Adler J, Back T, et~al., {AlphaFold 2
  presentation}; 2020.
\newblock [Online; accessed 2-May-2021].
\newblock
  \url{https://predictioncenter.org/casp14/doc/presentations/2020_12_01_TS_predictor_AlphaFold2.pdf}.

\bibitem[{Kandathil et~al.(2019)Kandathil, Shaun M and Greener, Joe G and
  Jones, David T}]{kandathil2019recent}
\textbf{\color{eLifeMediumGrey} Kandathil SM}, Greener JG, Jones DT.
\newblock Recent developments in deep learning applied to protein structure
  prediction.
\newblock Proteins: Structure, Function, and Bioinformatics.  2019;
  87(12):1179--1189.

\bibitem[{Kassem et~al.(2018)Kassem, Maher M and Christoffersen, Lars B and
  Cavalli, Andrea and Lindorff-Larsen, Kresten}]{kassem2018enhancing}
\textbf{\color{eLifeMediumGrey} Kassem MM}, Christoffersen LB, Cavalli A,
  Lindorff-Larsen K.
\newblock Enhancing coevolution-based contact prediction by imposing structural
  self-consistency of the contacts.
\newblock Scientific reports.  2018; 8(1):1--10.

\bibitem[{Khan et~al.(2013)Khan, Waqasuddin and Duffy, Fergal and Pollastri,
  Gianluca and Shields, Denis C and Mooney, Catherine}]{khan2013predicting}
\textbf{\color{eLifeMediumGrey} Khan W}, Duffy F, Pollastri G, Shields DC,
  Mooney C.
\newblock Predicting binding within disordered protein regions to structurally
  characterised peptide-binding domains.
\newblock PLoS One.  2013; 8(9):e72838.

\bibitem[{Kohlhoff et~al.(2009)Kohlhoff, Kai J and Robustelli, Paul and
  Cavalli, Andrea and Salvatella, Xavier and Vendruscolo,
  Michele}]{kohlhoff2009fast}
\textbf{\color{eLifeMediumGrey} Kohlhoff KJ}, Robustelli P, Cavalli A,
  Salvatella X, Vendruscolo M.
\newblock Fast and accurate predictions of protein NMR chemical shifts from
  interatomic distances.
\newblock Journal of the American Chemical Society.  2009;
  131(39):13894--13895.

\bibitem[{Koren et~al.(2018)Koren, Itay and Timms, Richard T and Kula, Tomasz
  and Xu, Qikai and Li, Mamie Z and Elledge, Stephen J}]{koren2018eukaryotic}
\textbf{\color{eLifeMediumGrey} Koren I}, Timms RT, Kula T, Xu Q, Li MZ,
  Elledge SJ.
\newblock The eukaryotic proteome is shaped by E3 ubiquitin ligases targeting
  C-terminal degrons.
\newblock Cell.  2018; 173(7):1622--1635.

\bibitem[{Kryshtafovych et~al.(2019)Kryshtafovych, Andriy and Schwede, Torsten
  and Topf, Maya and Fidelis, Krzysztof and Moult,
  John}]{kryshtafovych2019critical}
\textbf{\color{eLifeMediumGrey} Kryshtafovych A}, Schwede T, Topf M, Fidelis K,
  Moult J.
\newblock Critical assessment of methods of protein structure prediction
  (CASP)—Round XIII.
\newblock Proteins: Structure, Function, and Bioinformatics.  2019;
  87(12):1011--1020.

\bibitem[{Kumar et~al.(2020)Kumar, Manjeet and Gouw, Marc and Michael, Sushama
  and S{\'a}mano-S{\'a}nchez, Hugo and Pancsa, Rita and Glavina, Juliana and
  Diakogianni, Athina and Valverde, Jes{\'u}s Alvarado and Bukirova, Dayana and
  {\v{C}}aly{\v{s}}eva, Jelena and others}]{kumar2020elm}
\textbf{\color{eLifeMediumGrey} Kumar M}, Gouw M, Michael S,
  S{\'a}mano-S{\'a}nchez H, Pancsa R, Glavina J, Diakogianni A, Valverde JA,
  Bukirova D, {\v{C}}aly{\v{s}}eva J, et~al.
\newblock ELM—the eukaryotic linear motif resource in 2020.
\newblock Nucleic acids research.  2020; 48(D1):D296--D306.

\bibitem[{Kundu and Backofen(2014)Kundu, Kousik and Backofen,
  Rolf}]{kundu2014cluster}
\textbf{\color{eLifeMediumGrey} Kundu K}, Backofen R.
\newblock Cluster based prediction of PDZ-peptide interactions.
\newblock BMC genomics.  2014; 15(1):1--11.

\bibitem[{Laine et~al.(2021)Elodie Laine and Stephan Eismann and Arne Elofsson
  and Sergei Grudinin}]{laine2021protein}
\textbf{\color{eLifeMediumGrey} Laine E}, Eismann S, Elofsson A, Grudinin S.
\newblock Protein sequence-to-structure learning: Is this the end(-to-end
  revolution)?
\newblock arXiv.  2021; p. 2105.07407.

\bibitem[{Lapedes et~al.(2012)Lapedes, Alan and Giraud, Bertrand and Jarzynski,
  Christopher}]{lapedes2012using}
\textbf{\color{eLifeMediumGrey} Lapedes A}, Giraud B, Jarzynski C.
\newblock Using sequence alignments to predict protein structure and stability
  with high accuracy.
\newblock arXiv preprint arXiv:12072484.  2012; .

\bibitem[{Larsen et~al.(2020)Larsen, Andreas Haahr and Wang, Yong and Bottaro,
  Sandro and Grudinin, Sergei and Arleth, Lise and Lindorff-Larsen,
  Kresten}]{larsen2020combining}
\textbf{\color{eLifeMediumGrey} Larsen AH}, Wang Y, Bottaro S, Grudinin S,
  Arleth L, Lindorff-Larsen K.
\newblock Combining molecular dynamics simulations with small-angle X-ray and
  neutron scattering data to study multi-domain proteins in solution.
\newblock PLoS computational biology.  2020; 16(4):e1007870.

\bibitem[{Latham and Zhang(2019)Latham, Andrew P and Zhang,
  Bin}]{latham2019maximum}
\textbf{\color{eLifeMediumGrey} Latham AP}, Zhang B.
\newblock Maximum entropy optimized force field for intrinsically disordered
  proteins.
\newblock Journal of chemical theory and computation.  2019; 16(1):773--781.

\bibitem[{Lazar et~al.(2021)Lazar, Tamas and Mart{\'\i}nez-P{\'e}rez, Elizabeth
  and Quaglia, Federica and Hatos, Andr{\'a}s and Chemes, Luc{\'\i}a B and
  Iserte, Javier A and M{\'e}ndez, Nicol{\'a}s A and Garrone, Nicol{\'a}s A and
  Salda{\~n}o, Tadeo E and Marchetti, Julia and others}]{lazar2021ped}
\textbf{\color{eLifeMediumGrey} Lazar T}, Mart{\'\i}nez-P{\'e}rez E, Quaglia F,
  Hatos A, Chemes LB, Iserte JA, M{\'e}ndez NA, Garrone NA, Salda{\~n}o TE,
  Marchetti J, et~al.
\newblock PED in 2021: a major update of the protein ensemble database for
  intrinsically disordered proteins.
\newblock Nucleic Acids Research.  2021; 49(D1):D404--D411.

\bibitem[{van~der Lee et~al.(2014)van der Lee, Robin and Lang, Benjamin and
  Kruse, Kai and Gsponer, J{\"o}rg and de Groot, Natalia S{\'a}nchez and
  Huynen, Martijn A and Matouschek, Andreas and Fuxreiter, Monika and Babu, M
  Madan}]{van2014intrinsically}
\textbf{\color{eLifeMediumGrey} van~der Lee R}, Lang B, Kruse K, Gsponer J,
  de~Groot NS, Huynen MA, Matouschek A, Fuxreiter M, Babu MM.
\newblock Intrinsically disordered segments affect protein half-life in the
  cell and during evolution.
\newblock Cell reports.  2014; 8(6):1832--1844.

\bibitem[{Lee et~al.(2008)Lee, Yang David and Wang, Jun and Stubbe, JoAnne and
  Elledge, Stephen J}]{lee2008dif1}
\textbf{\color{eLifeMediumGrey} Lee YD}, Wang J, Stubbe J, Elledge SJ.
\newblock Dif1 is a DNA-damage-regulated facilitator of nuclear import for
  ribonucleotide reductase.
\newblock Molecular cell.  2008; 32(1):70--80.

\bibitem[{Li and Br{\"u}schweiler(2010)Li, Da-Wei and Br{\"u}schweiler,
  Rafael}]{li2010nmr}
\textbf{\color{eLifeMediumGrey} Li DW}, Br{\"u}schweiler R.
\newblock NMR-based protein potentials.
\newblock Angewandte Chemie International Edition.  2010; 49(38):6778--6780.

\bibitem[{Li and Br{\"u}schweiler(2012)Li, Da-Wei and Br{\"u}schweiler,
  Rafael}]{li2012ppm}
\textbf{\color{eLifeMediumGrey} Li DW}, Br{\"u}schweiler R.
\newblock PPM: a side-chain and backbone chemical shift predictor for the
  assessment of protein conformational ensembles.
\newblock Journal of biomolecular NMR.  2012; 54(3):257--265.

\bibitem[{Li et~al.(2020{\natexlab{a}})Li, Jie and Bennett, Kochise C and Liu,
  Yuchen and Martin, Michael V and Head-Gordon, Teresa}]{li2020accurate}
\textbf{\color{eLifeMediumGrey} Li J}, Bennett KC, Liu Y, Martin MV,
  Head-Gordon T.
\newblock Accurate prediction of chemical shifts for aqueous protein structure
  on “Real World” data.
\newblock Chemical Science.  2020; 11(12):3180--3191.

\bibitem[{Li et~al.(2017)Li, Jing and White, Jordan T and Saavedra, Harry and
  Wrabl, James O and Motlagh, Hesam N and Liu, Kaixian and Sowers, James and
  Schroer, Trina A and Thompson, E Brad and Hilser, Vincent
  J}]{li2017genetically}
\textbf{\color{eLifeMediumGrey} Li J}, White JT, Saavedra H, Wrabl JO, Motlagh
  HN, Liu K, Sowers J, Schroer TA, Thompson EB, Hilser VJ.
\newblock Genetically tunable frustration controls allostery in an
  intrinsically disordered transcription factor.
\newblock Elife.  2017; 6:e30688.

\bibitem[{Li et~al.(2012)Li, Pilong and Banjade, Sudeep and Cheng, Hui-Chun and
  Kim, Soyeon and Chen, Baoyu and Guo, Liang and Llaguno, Marc and
  Hollingsworth, Javoris V and King, David S and Banani, Salman F and
  others}]{li2012phase}
\textbf{\color{eLifeMediumGrey} Li P}, Banjade S, Cheng HC, Kim S, Chen B, Guo
  L, Llaguno M, Hollingsworth JV, King DS, Banani SF, et~al.
\newblock Phase transitions in the assembly of multivalent signalling proteins.
\newblock Nature.  2012; 483(7389):336--340.

\bibitem[{Li et~al.(2020{\natexlab{b}})Li, Qian and Peng, Xiaojun and Li,
  Yuanqing and Tang, Wenqin and Zhu, Jia’an and Huang, Jing and Qi, Yifei and
  Zhang, Zhuqing}]{li2020llpsdb}
\textbf{\color{eLifeMediumGrey} Li Q}, Peng X, Li Y, Tang W, Zhu J, Huang J, Qi
  Y, Zhang Z.
\newblock LLPSDB: a database of proteins undergoing liquid--liquid phase
  separation in vitro.
\newblock Nucleic acids research.  2020; 48(D1):D320--D327.

\bibitem[{Li et~al.(2020{\natexlab{c}})Li, Qian and Wang, Xi and Dou, Zhihui
  and Yang, Weishan and Huang, Beifang and Lou, Jizhong and Zhang,
  Zhuqing}]{li2020protein}
\textbf{\color{eLifeMediumGrey} Li Q}, Wang X, Dou Z, Yang W, Huang B, Lou J,
  Zhang Z.
\newblock Protein Databases Related to Liquid--Liquid Phase Separation.
\newblock International Journal of Molecular Sciences.  2020; 21(18):6796.

\bibitem[{Li et~al.(2019)Li, Yongsheng and Zhang, Yunpeng and Li, Xia and Yi,
  Song and Xu, Juan}]{li2019gain}
\textbf{\color{eLifeMediumGrey} Li Y}, Zhang Y, Li X, Yi S, Xu J.
\newblock Gain-of-function mutations: an emerging advantage for cancer biology.
\newblock Trends in biochemical sciences.  2019; 44(8):659--674.

\bibitem[{Light et~al.(2013)Light, Sara and Sagit, Rauan and Sachenkova, Oxana
  and Ekman, Diana and Elofsson, Arne}]{light2013protein}
\textbf{\color{eLifeMediumGrey} Light S}, Sagit R, Sachenkova O, Ekman D,
  Elofsson A.
\newblock Protein expansion is primarily due to indels in intrinsically
  disordered regions.
\newblock Molecular biology and evolution.  2013; 30(12):2645--2653.

\bibitem[{Lin and Chan(2017)Lin, Yi-Hsuan and Chan, Hue Sun}]{lin2017phase}
\textbf{\color{eLifeMediumGrey} Lin YH}, Chan HS.
\newblock Phase separation and single-chain compactness of charged disordered
  proteins are strongly correlated.
\newblock Biophysical Journal.  2017; 112(10):2043--2046.

\bibitem[{Lindorff-Larsen et~al.(2005)Lindorff-Larsen, Kresten and Best, Robert
  B and Vendruscolo, Michele}]{lindorff2005interpreting}
\textbf{\color{eLifeMediumGrey} Lindorff-Larsen K}, Best RB, Vendruscolo M.
\newblock Interpreting dynamically-averaged scalar couplings in proteins.
\newblock Journal of biomolecular NMR.  2005; 32(4):273--280.

\bibitem[{Lindorff-Larsen and Ferkinghoff-Borg(2009)Lindorff-Larsen, Kresten
  and Ferkinghoff-Borg, Jesper}]{lindorff2009similarity}
\textbf{\color{eLifeMediumGrey} Lindorff-Larsen K}, Ferkinghoff-Borg J.
\newblock Similarity measures for protein ensembles.
\newblock PloS one.  2009; 4(1):e4203.

\bibitem[{Livesey and Marsh(2020)Livesey, Benjamin J and Marsh, Joseph
  A}]{livesey2020using}
\textbf{\color{eLifeMediumGrey} Livesey BJ}, Marsh JA.
\newblock Using deep mutational scanning to benchmark variant effect predictors
  and identify disease mutations.
\newblock Molecular systems biology.  2020; 16(7):e9380.

\bibitem[{Lu et~al.(2018)Lu, Huasong and Yu, Dan and Hansen, Anders S and
  Ganguly, Sourav and Liu, Rongdiao and Heckert, Alec and Darzacq, Xavier and
  Zhou, Qiang}]{lu2018phase}
\textbf{\color{eLifeMediumGrey} Lu H}, Yu D, Hansen AS, Ganguly S, Liu R,
  Heckert A, Darzacq X, Zhou Q.
\newblock Phase-separation mechanism for C-terminal hyperphosphorylation of RNA
  polymerase II.
\newblock Nature.  2018; 558(7709):318--323.

\bibitem[{Madeira et~al.(2015)Madeira, F{\'a}bio and Tinti, Michele and
  Murugesan, Gavuthami and Berrett, Emily and Stafford, Margaret and Toth,
  Rachel and Cole, Christian and MacKintosh, Carol and Barton, Geoffrey
  J}]{madeira201514}
\textbf{\color{eLifeMediumGrey} Madeira F}, Tinti M, Murugesan G, Berrett E,
  Stafford M, Toth R, Cole C, MacKintosh C, Barton GJ.
\newblock 14-3-3-Pred: improved methods to predict 14-3-3-binding
  phosphopeptides.
\newblock Bioinformatics.  2015; 31(14):2276--2283.

\bibitem[{Marks et~al.(2011)Marks, Debora S and Colwell, Lucy J and Sheridan,
  Robert and Hopf, Thomas A and Pagnani, Andrea and Zecchina, Riccardo and
  Sander, Chris}]{marks2011protein}
\textbf{\color{eLifeMediumGrey} Marks DS}, Colwell LJ, Sheridan R, Hopf TA,
  Pagnani A, Zecchina R, Sander C.
\newblock Protein 3D structure computed from evolutionary sequence variation.
\newblock PloS one.  2011; 6(12):e28766.

\bibitem[{Marsh and Forman-Kay(2010)Marsh, Joseph A and Forman-Kay, Julie
  D}]{marsh2010sequence}
\textbf{\color{eLifeMediumGrey} Marsh JA}, Forman-Kay JD.
\newblock Sequence determinants of compaction in intrinsically disordered
  proteins.
\newblock Biophysical journal.  2010; 98(10):2383--2390.

\bibitem[{Martin et~al.(2020)Martin, Erik W and Holehouse, Alex S and Peran,
  Ivan and Farag, Mina and Incicco, J Jeremias and Bremer, Anne and Grace,
  Christy R and Soranno, Andrea and Pappu, Rohit V and Mittag,
  Tanja}]{martin2020valence}
\textbf{\color{eLifeMediumGrey} Martin EW}, Holehouse AS, Peran I, Farag M,
  Incicco JJ, Bremer A, Grace CR, Soranno A, Pappu RV, Mittag T.
\newblock Valence and patterning of aromatic residues determine the phase
  behavior of prion-like domains.
\newblock Science.  2020; 367(6478):694--699.

\bibitem[{Matreyek et~al.(2018)Matreyek, Kenneth A and Starita, Lea M and
  Stephany, Jason J and Martin, Beth and Chiasson, Melissa A and Gray, Vanessa
  E and Kircher, Martin and Khechaduri, Arineh and Dines, Jennifer N and Hause,
  Ronald J and others}]{matreyek2018multiplex}
\textbf{\color{eLifeMediumGrey} Matreyek KA}, Starita LM, Stephany JJ, Martin
  B, Chiasson MA, Gray VE, Kircher M, Khechaduri A, Dines JN, Hause RJ, et~al.
\newblock Multiplex assessment of protein variant abundance by massively
  parallel sequencing.
\newblock Nature genetics.  2018; 50(6):874--882.

\bibitem[{Meiler(2003)Meiler, Jens}]{meiler2003proshift}
\textbf{\color{eLifeMediumGrey} Meiler J}.
\newblock PROSHIFT: protein chemical shift prediction using artificial neural
  networks.
\newblock Journal of biomolecular NMR.  2003; 26(1):25--37.

\bibitem[{M{\'e}sz{\'a}ros et~al.(2020)M{\'e}sz{\'a}ros, B{\'a}lint and
  Erd{\H{o}}s, G{\'a}bor and Szab{\'o}, Be{\'a}ta and Sch{\'a}d, {\'E}va and
  Tantos, {\'A}gnes and Abukhairan, Rawan and Horv{\'a}th, Tam{\'a}s and
  Murvai, Nikoletta and Kov{\'a}cs, Orsolya P and Kov{\'a}cs, M{\'a}rton and
  others}]{meszaros2020phasepro}
\textbf{\color{eLifeMediumGrey} M{\'e}sz{\'a}ros B}, Erd{\H{o}}s G, Szab{\'o}
  B, Sch{\'a}d {\'E}, Tantos {\'A}, Abukhairan R, Horv{\'a}th T, Murvai N,
  Kov{\'a}cs OP, Kov{\'a}cs M, et~al.
\newblock PhaSePro: the database of proteins driving liquid--liquid phase
  separation.
\newblock Nucleic acids research.  2020; 48(D1):D360--D367.

\bibitem[{M{\'e}sz{\'a}ros et~al.(2021)M{\'e}sz{\'a}ros, B{\'a}lint and
  Hajdu-Solt{\'e}sz, Borb{\'a}la and Zeke, Andr{\'a}s and Doszt{\'a}nyi,
  Zsuzsanna}]{meszaros2021mutations}
\textbf{\color{eLifeMediumGrey} M{\'e}sz{\'a}ros B}, Hajdu-Solt{\'e}sz B, Zeke
  A, Doszt{\'a}nyi Z.
\newblock Mutations of Intrinsically Disordered Protein Regions Can Drive
  Cancer but Lack Therapeutic Strategies.
\newblock Biomolecules.  2021; 11(3):381.

\bibitem[{Meyer et~al.(2018)Meyer, Katrina and Kirchner, Marieluise and Uyar,
  Bora and Cheng, Jing-Yuan and Russo, Giulia and Hernandez-Miranda, Luis R and
  Szymborska, Anna and Zauber, Henrik and Rudolph, Ina-Maria and Willnow,
  Thomas E and others}]{meyer2018mutations}
\textbf{\color{eLifeMediumGrey} Meyer K}, Kirchner M, Uyar B, Cheng JY, Russo
  G, Hernandez-Miranda LR, Szymborska A, Zauber H, Rudolph IM, Willnow TE,
  et~al.
\newblock Mutations in disordered regions can cause disease by creating
  dileucine motifs.
\newblock Cell.  2018; 175(1):239--253.

\bibitem[{van Mierlo et~al.(2021)van Mierlo, Guido and Jansen, Jurriaan RG and
  Wang, Jie and Poser, Ina and van Heeringen, Simon J and Vermeulen,
  Michiel}]{van2021predicting}
\textbf{\color{eLifeMediumGrey} van Mierlo G}, Jansen JR, Wang J, Poser I, van
  Heeringen SJ, Vermeulen M.
\newblock Predicting protein condensate formation using machine learning.
\newblock Cell Reports.  2021; 34(5):108705.

\bibitem[{Milles et~al.(2018)Milles, Sigrid and Jensen, Malene Ringkj{\o}bing
  and Lazert, Carine and Guseva, Serafima and Ivashchenko, Stefaniia and
  Communie, Guillaume and Maurin, Damien and Gerlier, Denis and Ruigrok, Rob WH
  and Blackledge, Martin}]{milles2018ultraweak}
\textbf{\color{eLifeMediumGrey} Milles S}, Jensen MR, Lazert C, Guseva S,
  Ivashchenko S, Communie G, Maurin D, Gerlier D, Ruigrok RW, Blackledge M.
\newblock An ultraweak interaction in the intrinsically disordered replication
  machinery is essential for measles virus function.
\newblock Science advances.  2018; 4(8):eaat7778.

\bibitem[{Mittag and Forman-Kay(2007)Mittag, Tanja and Forman-Kay, Julie
  D}]{mittag2007atomic}
\textbf{\color{eLifeMediumGrey} Mittag T}, Forman-Kay JD.
\newblock Atomic-level characterization of disordered protein ensembles.
\newblock Current opinion in structural biology.  2007; 17(1):3--14.

\bibitem[{Monahan et~al.(2017)Monahan, Zachary and Ryan, Veronica H and Janke,
  Abigail M and Burke, Kathleen A and Rhoads, Shannon N and Zerze, G{\"u}l H
  and O'Meally, Robert and Dignon, Gregory L and Conicella, Alexander E and
  Zheng, Wenwei and others}]{monahan2017phosphorylation}
\textbf{\color{eLifeMediumGrey} Monahan Z}, Ryan VH, Janke AM, Burke KA, Rhoads
  SN, Zerze GH, O'Meally R, Dignon GL, Conicella AE, Zheng W, et~al.
\newblock Phosphorylation of the FUS low-complexity domain disrupts phase
  separation, aggregation, and toxicity.
\newblock The EMBO journal.  2017; 36(20):2951--2967.

\bibitem[{Morcos et~al.(2011)Morcos, Faruck and Pagnani, Andrea and Lunt, Bryan
  and Bertolino, Arianna and Marks, Debora S and Sander, Chris and Zecchina,
  Riccardo and Onuchic, Jos{\'e} N and Hwa, Terence and Weigt,
  Martin}]{morcos2011direct}
\textbf{\color{eLifeMediumGrey} Morcos F}, Pagnani A, Lunt B, Bertolino A,
  Marks DS, Sander C, Zecchina R, Onuchic JN, Hwa T, Weigt M.
\newblock Direct-coupling analysis of residue coevolution captures native
  contacts across many protein families.
\newblock Proceedings of the National Academy of Sciences.  2011;
  108(49):E1293--E1301.

\bibitem[{Moult et~al.(1995)Moult, John and Pedersen, Jan T and Judson, Richard
  and Fidelis, Krzysztof}]{moult1995large}
\textbf{\color{eLifeMediumGrey} Moult J}, Pedersen JT, Judson R, Fidelis K.
\newblock A large-scale experiment to assess protein structure prediction
  methods.
\newblock Proteins.  1995; 23:ii--iv.

\bibitem[{Mu et~al.(2021)Mu, Junxi and Liu, Hao and Zhang, Jian and Luo, Ray
  and Chen, Hai-Feng}]{mu2021recent}
\textbf{\color{eLifeMediumGrey} Mu J}, Liu H, Zhang J, Luo R, Chen HF.
\newblock Recent Force Field Strategies for Intrinsically Disordered Proteins.
\newblock Journal of Chemical Information and Modeling.  2021;
  61(3):1037--1047.

\bibitem[{Nagy et~al.(2019)Nagy, Gabor and Igaev, Maxim and Jones, Nykola C and
  Hoffmann, S{\o}ren V and Grubm{\"u}ller, Helmut}]{nagy2019sesca}
\textbf{\color{eLifeMediumGrey} Nagy G}, Igaev M, Jones NC, Hoffmann SV,
  Grubm{\"u}ller H.
\newblock SESCA: predicting circular dichroism spectra from protein molecular
  structures.
\newblock Journal of chemical theory and computation.  2019; 15(9):5087--5102.

\bibitem[{Necci et~al.(2021)Necci, Marco and Piovesan, Damiano and Tosatto,
  Silvio CE}]{necci2021critical}
\textbf{\color{eLifeMediumGrey} Necci M}, Piovesan D, Tosatto SC.
\newblock Critical assessment of protein intrinsic disorder prediction.
\newblock Nature Methods.  2021; p. 1--10.

\bibitem[{Neduva et~al.(2005)Neduva, Victor and Linding, Rune and Su-Angrand,
  Isabelle and Stark, Alexander and De Masi, Federico and Gibson, Toby J and
  Lewis, Joe and Serrano, Luis and Russell, Robert B}]{neduva2005systematic}
\textbf{\color{eLifeMediumGrey} Neduva V}, Linding R, Su-Angrand I, Stark A,
  De~Masi F, Gibson TJ, Lewis J, Serrano L, Russell RB.
\newblock Systematic discovery of new recognition peptides mediating protein
  interaction networks.
\newblock PLoS Biol.  2005; 3(12):e405.

\bibitem[{Neher(1994)Neher, Erwin}]{neher1994frequent}
\textbf{\color{eLifeMediumGrey} Neher E}.
\newblock How frequent are correlated changes in families of protein sequences?
\newblock Proceedings of the National Academy of Sciences.  1994;
  91(1):98--102.

\bibitem[{Newberry et~al.(2020{\natexlab{a}})Newberry, Robert W and Arhar,
  Taylor and Costello, Jean and Hartoularos, George C and Maxwell, Alison M and
  Naing, Zun Zar Chi and Pittman, Maureen and Reddy, Nishith R and Schwarz,
  Daniel MC and Wassarman, Douglas R and others}]{newberry2020robust}
\textbf{\color{eLifeMediumGrey} Newberry RW}, Arhar T, Costello J, Hartoularos
  GC, Maxwell AM, Naing ZZC, Pittman M, Reddy NR, Schwarz DM, Wassarman DR,
  et~al.
\newblock Robust Sequence Determinants of $\alpha$-Synuclein Toxicity in Yeast
  Implicate Membrane Binding.
\newblock ACS Chemical Biology.  2020; 15(8):2137--2153.

\bibitem[{Newberry et~al.(2020{\natexlab{b}})Newberry, Robert W and Leong,
  Jaime T and Chow, Eric D and Kampmann, Martin and DeGrado, William
  F}]{newberry2020deep}
\textbf{\color{eLifeMediumGrey} Newberry RW}, Leong JT, Chow ED, Kampmann M,
  DeGrado WF.
\newblock Deep mutational scanning reveals the structural basis for
  $\alpha$-synuclein activity.
\newblock Nature chemical biology.  2020; 16(6):653--659.

\bibitem[{Nielsen and Mulder(2019)Nielsen, Jakob T and Mulder, Frans
  AA}]{nielsen2019quality}
\textbf{\color{eLifeMediumGrey} Nielsen JT}, Mulder FA.
\newblock Quality and bias of protein disorder predictors.
\newblock Scientific reports.  2019; 9(1):1--11.

\bibitem[{Ning et~al.(2020)Ning, Wanshan and Guo, Yaping and Lin, Shaofeng and
  Mei, Bin and Wu, Yu and Jiang, Peiran and Tan, Xiaodan and Zhang, Weizhi and
  Chen, Guowei and Peng, Di and others}]{ning2020drllps}
\textbf{\color{eLifeMediumGrey} Ning W}, Guo Y, Lin S, Mei B, Wu Y, Jiang P,
  Tan X, Zhang W, Chen G, Peng D, et~al.
\newblock DrLLPS: a data resource of liquid--liquid phase separation in
  eukaryotes.
\newblock Nucleic acids research.  2020; 48(D1):D288--D295.

\bibitem[{Njo et~al.(1995)Njo, Swie Lan and van Gunsteren, Wilfred F and
  M{\"u}ller-Plathe, Florian}]{njo1995determination}
\textbf{\color{eLifeMediumGrey} Njo SL}, van Gunsteren WF, M{\"u}ller-Plathe F.
\newblock Determination of force field parameters for molecular simulation by
  molecular simulation: An application of the weak-coupling method.
\newblock The Journal of chemical physics.  1995; 102(15):6199--6207.

\bibitem[{Noid(2013)Noid, William George}]{noid2013perspective}
\textbf{\color{eLifeMediumGrey} Noid WG}.
\newblock Perspective: Coarse-grained models for biomolecular systems.
\newblock The Journal of chemical physics.  2013; 139(9):09B201\_1.

\bibitem[{Norgaard et~al.(2008)Norgaard, Anders B and Ferkinghoff-Borg, Jesper
  and Lindorff-Larsen, Kresten}]{norgaard2008experimental}
\textbf{\color{eLifeMediumGrey} Norgaard AB}, Ferkinghoff-Borg J,
  Lindorff-Larsen K.
\newblock Experimental parameterization of an energy function for the
  simulation of unfolded proteins.
\newblock Biophysical journal.  2008; 94(1):182--192.

\bibitem[{Norrby and Liljefors(1998)Norrby, Per-Ola and Liljefors,
  Tommy}]{norrby1998automated}
\textbf{\color{eLifeMediumGrey} Norrby PO}, Liljefors T.
\newblock Automated molecular mechanics parameterization with simultaneous
  utilization of experimental and quantum mechanical data.
\newblock Journal of computational chemistry.  1998; 19(10):1146--1166.

\bibitem[{Nott et~al.(2015)Nott, Timothy J and Petsalaki, Evangelia and Farber,
  Patrick and Jervis, Dylan and Fussner, Eden and Plochowietz, Anne and Craggs,
  Timothy D and Bazett-Jones, David P and Pawson, Tony and Forman-Kay, Julie D
  and others}]{nott2015phase}
\textbf{\color{eLifeMediumGrey} Nott TJ}, Petsalaki E, Farber P, Jervis D,
  Fussner E, Plochowietz A, Craggs TD, Bazett-Jones DP, Pawson T, Forman-Kay
  JD, et~al.
\newblock Phase transition of a disordered nuage protein generates
  environmentally responsive membraneless organelles.
\newblock Molecular cell.  2015; 57(5):936--947.

\bibitem[{Ofer et~al.(2021)Ofer, Dan and Brandes, Nadav and Linial,
  Michal}]{ofer2021language}
\textbf{\color{eLifeMediumGrey} Ofer D}, Brandes N, Linial M.
\newblock The language of proteins: NLP, machine learning \& protein sequences.
\newblock Computational and Structural Biotechnology Journal.  2021; .

\bibitem[{Oldfield et~al.(2008)Oldfield, Christopher J and Meng, Jingwei and
  Yang, Jack Y and Yang, Mary Qu and Uversky, Vladimir N and Dunker, A
  Keith}]{oldfield2008flexible}
\textbf{\color{eLifeMediumGrey} Oldfield CJ}, Meng J, Yang JY, Yang MQ, Uversky
  VN, Dunker AK.
\newblock Flexible nets: disorder and induced fit in the associations of p53
  and 14-3-3 with their partners.
\newblock BMC genomics.  2008; 9(1):1--20.

\bibitem[{Olsen et~al.(2017)Olsen, Johan G and Teilum, Kaare and Kragelund,
  Birthe B}]{olsen2017behaviour}
\textbf{\color{eLifeMediumGrey} Olsen JG}, Teilum K, Kragelund BB.
\newblock Behaviour of intrinsically disordered proteins in protein--protein
  complexes with an emphasis on fuzziness.
\newblock Cellular and Molecular Life Sciences.  2017; 74(17):3175--3183.

\bibitem[{Orioli et~al.(2020)Orioli, Simone and Larsen, Andreas Haahr and
  Bottaro, Sandro and Lindorff-Larsen, Kresten}]{orioli2020learn}
\textbf{\color{eLifeMediumGrey} Orioli S}, Larsen AH, Bottaro S,
  Lindorff-Larsen K.
\newblock How to learn from inconsistencies: Integrating molecular simulations
  with experimental data.
\newblock Progress in molecular biology and translational science.  2020;
  170:123--176.

\bibitem[{O'Shea et~al.(2017)O'Shea, Charlotte and Staby, Lasse and Bendsen,
  Sidsel Krogh and Tidemand, Frederik Gr{\o}nb{\ae}k and Redsted, Andreas and
  Willemo{\"e}s, Martin and Kragelund, Birthe B and Skriver,
  Karen}]{o2017structures}
\textbf{\color{eLifeMediumGrey} O'Shea C}, Staby L, Bendsen SK, Tidemand FG,
  Redsted A, Willemo{\"e}s M, Kragelund BB, Skriver K.
\newblock Structures and short linear motif of disordered transcription factor
  regions provide clues to the interactome of the cellular hub protein
  radical-induced cell death1.
\newblock Journal of Biological Chemistry.  2017; 292(2):512--527.

\bibitem[{Palopoli et~al.(2018)Palopoli, Nicol{\'a}s and Gonz{\'a}lez Foutel,
  Nicol{\'a}s S and Gibson, Toby J and Chemes, Luc{\'\i}a
  B}]{palopoli2018short}
\textbf{\color{eLifeMediumGrey} Palopoli N}, Gonz{\'a}lez~Foutel NS, Gibson TJ,
  Chemes LB.
\newblock Short linear motif core and flanking regions modulate retinoblastoma
  protein binding affinity and specificity.
\newblock Protein Engineering, Design and Selection.  2018; 31(3):69--77.

\bibitem[{Panagiotopoulos et~al.(1998)Panagiotopoulos, Athanassios Z and Wong,
  Vicky and Floriano, M Antonio}]{panagiotopoulos1998phase}
\textbf{\color{eLifeMediumGrey} Panagiotopoulos AZ}, Wong V, Floriano MA.
\newblock Phase equilibria of lattice polymers from histogram reweighting Monte
  Carlo simulations.
\newblock Macromolecules.  1998; 31(3):912--918.

\bibitem[{Pawar et~al.(2005)Pawar, Amol P and Dubay, Kateri F and Zurdo, Jesus
  and Chiti, Fabrizio and Vendruscolo, Michele and Dobson, Christopher
  M}]{pawar2005prediction}
\textbf{\color{eLifeMediumGrey} Pawar AP}, Dubay KF, Zurdo J, Chiti F,
  Vendruscolo M, Dobson CM.
\newblock Prediction of “aggregation-prone” and
  “aggregation-susceptible” regions in proteins associated with
  neurodegenerative diseases.
\newblock Journal of molecular biology.  2005; 350(2):379--392.

\bibitem[{Peran and Mittag(2020)Peran, Ivan and Mittag,
  Tanja}]{peran2020molecular}
\textbf{\color{eLifeMediumGrey} Peran I}, Mittag T.
\newblock Molecular structure in biomolecular condensates.
\newblock Current opinion in structural biology.  2020; 60:17--26.

\bibitem[{Perham(1975)Perham, Richard Nelson}]{perham1975self}
\textbf{\color{eLifeMediumGrey} Perham RN}.
\newblock Self-assembly of biological macromolecules.
\newblock Philosophical Transactions of the Royal Society of London B,
  Biological Sciences.  1975; 272(915):123--136.

\bibitem[{Pesce and Lindorff-Larsen(2021)Pesce, Francesco and Lindorff-Larsen,
  Kresten}]{pesce2021refining}
\textbf{\color{eLifeMediumGrey} Pesce F}, Lindorff-Larsen K.
\newblock Refining conformational ensembles of flexible proteins against
  small-angle X-ray scattering data.
\newblock bioRxiv.  2021; \href{10.1101/2021.05.29.446281}{\doiprefix
  \detokenize{10.1101/2021.05.29.446281}}.

\bibitem[{Piana et~al.(2011)Piana, Stefano and Lindorff-Larsen, Kresten and
  Shaw, David E}]{piana2011robust}
\textbf{\color{eLifeMediumGrey} Piana S}, Lindorff-Larsen K, Shaw DE.
\newblock How robust are protein folding simulations with respect to force
  field parameterization?
\newblock Biophysical journal.  2011; 100(9):L47--L49.

\bibitem[{Piovesan et~al.(2021)Piovesan, Damiano and Necci, Marco and Escobedo,
  Nahuel and Monzon, Alexander Miguel and Hatos, Andr{\'a}s and
  Mi{\v{c}}eti{\'c}, Ivan and Quaglia, Federica and Paladin, Lisanna and
  Ramasamy, Pathmanaban and Doszt{\'a}nyi, Zsuzsanna and
  others}]{piovesan2021mobidb}
\textbf{\color{eLifeMediumGrey} Piovesan D}, Necci M, Escobedo N, Monzon AM,
  Hatos A, Mi{\v{c}}eti{\'c} I, Quaglia F, Paladin L, Ramasamy P, Doszt{\'a}nyi
  Z, et~al.
\newblock MobiDB: intrinsically disordered proteins in 2021.
\newblock Nucleic Acids Research.  2021; 49(D1):D361--D367.

\bibitem[{Plewczy{\'n}ski et~al.(2005)Plewczy{\'n}ski, DARIUSZ and Tkacz,
  Adrian and Godzik, Adam and Rychlewski, Leszek}]{plewczynski2005support}
\textbf{\color{eLifeMediumGrey} Plewczy{\'n}ski D}, Tkacz A, Godzik A,
  Rychlewski L.
\newblock A support vector machine approach to the identification of
  phosphorylation sites.
\newblock Cell Mol Biol Lett.  2005; 10(1):73--89.

\bibitem[{Prestel et~al.(2019)Prestel, Andreas and Wichmann, Nanna and Martins,
  Joao M and Marabini, Riccardo and Kassem, Noah and Broendum, Sebastian S and
  Otterlei, Marit and Nielsen, Olaf and Willemo{\"e}s, Martin and Ploug,
  Michael and others}]{prestel2019pcna}
\textbf{\color{eLifeMediumGrey} Prestel A}, Wichmann N, Martins JM, Marabini R,
  Kassem N, Broendum SS, Otterlei M, Nielsen O, Willemo{\"e}s M, Ploug M,
  et~al.
\newblock The PCNA interaction motifs revisited: thinking outside the PIP-box.
\newblock Cellular and Molecular Life Sciences.  2019; 76(24):4923--4943.

\bibitem[{Priti{\v{s}}anac et~al.(2019)Priti{\v{s}}anac, Iva and Vernon, Robert
  M and Moses, Alan M and Forman Kay, Julie D}]{pritivsanac2019entropy}
\textbf{\color{eLifeMediumGrey} Priti{\v{s}}anac I}, Vernon RM, Moses AM,
  Forman~Kay JD.
\newblock Entropy and information within intrinsically disordered protein
  regions.
\newblock Entropy.  2019; 21(7):662.

\bibitem[{Raimondi et~al.(2021)Raimondi, Daniele and Orlando, Gabriele and
  Michiels, Emiel and Pakravan, Donya and Bratek-Skicki, Anna and Van Den
  Bosch, Ludo and Moreau, Yves and Rousseau, Frederic and Schymkowitz,
  Joost}]{raimondi2021silico}
\textbf{\color{eLifeMediumGrey} Raimondi D}, Orlando G, Michiels E, Pakravan D,
  Bratek-Skicki A, Van Den~Bosch L, Moreau Y, Rousseau F, Schymkowitz J.
\newblock In-silico prediction of in-vitro protein liquid-liquid phase
  separation experiments outcomes with multi-head neural attention.
\newblock Bioinformatics.  2021; .

\bibitem[{Rao et~al.(2019)Rao, Roshan and Bhattacharya, Nicholas and Thomas,
  Neil and Duan, Yan and Chen, Xi and Canny, John and Abbeel, Pieter and Song,
  Yun S}]{rao2019evaluating}
\textbf{\color{eLifeMediumGrey} Rao R}, Bhattacharya N, Thomas N, Duan Y, Chen
  X, Canny J, Abbeel P, Song YS.
\newblock Evaluating protein transfer learning with tape.
\newblock Advances in Neural Information Processing Systems.  2019; 32:9689.

\bibitem[{Rao et~al.(2021)Rao, Roshan and Liu, Jason and Verkuil, Robert and
  Meier, Joshua and Canny, John F and Abbeel, Pieter and Sercu, Tom and Rives,
  Alexander}]{rao2021msa}
\textbf{\color{eLifeMediumGrey} Rao R}, Liu J, Verkuil R, Meier J, Canny JF,
  Abbeel P, Sercu T, Rives A.
\newblock Msa transformer.
\newblock bioRxiv.  2021; .

\bibitem[{Ravarani et~al.(2018)Ravarani, Charles NJ and Erkina, Tamara Y and De
  Baets, Greet and Dudman, Daniel C and Erkine, Alexandre M and Babu, M
  Madan}]{ravarani2018high}
\textbf{\color{eLifeMediumGrey} Ravarani CN}, Erkina TY, De~Baets G, Dudman DC,
  Erkine AM, Babu MM.
\newblock High-throughput discovery of functional disordered regions:
  investigation of transactivation domains.
\newblock Molecular systems biology.  2018; 14(5):e8190.

\bibitem[{Riback et~al.(2017)Riback, Joshua A and Bowman, Micayla A and
  Zmyslowski, Adam M and Knoverek, Catherine R and Jumper, John M and Hinshaw,
  James R and Kaye, Emily B and Freed, Karl F and Clark, Patricia L and
  Sosnick, Tobin R}]{riback2017innovative}
\textbf{\color{eLifeMediumGrey} Riback JA}, Bowman MA, Zmyslowski AM, Knoverek
  CR, Jumper JM, Hinshaw JR, Kaye EB, Freed KF, Clark PL, Sosnick TR.
\newblock Innovative scattering analysis shows that hydrophobic disordered
  proteins are expanded in water.
\newblock Science.  2017; 358(6360):238--241.

\bibitem[{Rieping et~al.(2005)Rieping, Wolfgang and Habeck, Michael and Nilges,
  Michael}]{rieping2005inferential}
\textbf{\color{eLifeMediumGrey} Rieping W}, Habeck M, Nilges M.
\newblock Inferential structure determination.
\newblock Science.  2005; 309(5732):303--306.

\bibitem[{Riesselman et~al.(2018)Riesselman, Adam J and Ingraham, John B and
  Marks, Debora S}]{riesselman2018deep}
\textbf{\color{eLifeMediumGrey} Riesselman AJ}, Ingraham JB, Marks DS.
\newblock Deep generative models of genetic variation capture the effects of
  mutations.
\newblock Nature methods.  2018; 15(10):816--822.

\bibitem[{Robustelli et~al.(2018)Robustelli, Paul and Piana, Stefano and Shaw,
  David E}]{robustelli2018developing}
\textbf{\color{eLifeMediumGrey} Robustelli P}, Piana S, Shaw DE.
\newblock Developing a molecular dynamics force field for both folded and
  disordered protein states.
\newblock Proceedings of the National Academy of Sciences.  2018;
  115(21):E4758--E4766.

\bibitem[{Robustelli et~al.(2020)Robustelli, Paul and Piana, Stefano and Shaw,
  David E}]{robustelli2020mechanism}
\textbf{\color{eLifeMediumGrey} Robustelli P}, Piana S, Shaw DE.
\newblock Mechanism of coupled folding-upon-binding of an intrinsically
  disordered protein.
\newblock Journal of the American Chemical Society.  2020;
  142(25):11092--11101.

\bibitem[{Rogers et~al.(2018)Rogers, Joseph M and Passioura, Toby and Suga,
  Hiroaki}]{rogers2018nonproteinogenic}
\textbf{\color{eLifeMediumGrey} Rogers JM}, Passioura T, Suga H.
\newblock Nonproteinogenic deep mutational scanning of linear and cyclic
  peptides.
\newblock Proceedings of the National Academy of Sciences.  2018;
  115(43):10959--10964.

\bibitem[{Rogers et~al.(2014)Rogers, Joseph M and Wong, Chi T and Clarke,
  Jane}]{rogers2014coupled}
\textbf{\color{eLifeMediumGrey} Rogers JM}, Wong CT, Clarke J.
\newblock Coupled folding and binding of the disordered protein PUMA does not
  require particular residual structure.
\newblock Journal of the American Chemical Society.  2014; 136(14):5197--5200.

\bibitem[{Ronan et~al.(2020)Ronan, Tom and Garnett, Roman and Naegle, Kristen
  M}]{ronan2020new}
\textbf{\color{eLifeMediumGrey} Ronan T}, Garnett R, Naegle KM.
\newblock New analysis pipeline for high-throughput domain--peptide affinity
  experiments improves SH2 interaction data.
\newblock Journal of Biological Chemistry.  2020; 295(32):11346--11363.

\bibitem[{Rozen et~al.(2015)Rozen, Shelly and F{\"u}zesi-Levi, Maria G and
  Ben-Nissan, Gili and Mizrachi, Limor and Gabashvili, Alexandra and Levin,
  Yishai and Ben-Dor, Shifra and Eisenstein, Miriam and Sharon,
  Michal}]{rozen2015csnap}
\textbf{\color{eLifeMediumGrey} Rozen S}, F{\"u}zesi-Levi MG, Ben-Nissan G,
  Mizrachi L, Gabashvili A, Levin Y, Ben-Dor S, Eisenstein M, Sharon M.
\newblock CSNAP is a stoichiometric subunit of the COP9 signalosome.
\newblock Cell reports.  2015; 13(3):585--598.

\bibitem[{Ruff et~al.(2015)Ruff, Kiersten M and Harmon, Tyler S and Pappu,
  Rohit V}]{ruff2015camelot}
\textbf{\color{eLifeMediumGrey} Ruff KM}, Harmon TS, Pappu RV.
\newblock CAMELOT: A machine learning approach for coarse-grained simulations
  of aggregation of block-copolymeric protein sequences.
\newblock The Journal of chemical physics.  2015; 143(24):12B607\_1.

\bibitem[{Saar et~al.(2021)Saar, Kadi L and Morgunov, Alexey S and Qi, Runzhang
  and Arter, William E and Krainer, Georg and Knowles, Tuomas PJ and
  others}]{saar2021learning}
\textbf{\color{eLifeMediumGrey} Saar KL}, Morgunov AS, Qi R, Arter WE, Krainer
  G, Knowles TP, et~al.
\newblock Learning the molecular grammar of protein condensates from sequence
  determinants and embeddings.
\newblock Proceedings of the National Academy of Sciences.  2021; 118(15).

\bibitem[{Sahni et~al.(2015)Sahni, Nidhi and Yi, Song and Taipale, Mikko and
  Bass, Juan I Fuxman and Coulombe-Huntington, Jasmin and Yang, Fan and Peng,
  Jian and Weile, Jochen and Karras, Georgios I and Wang, Yang and
  others}]{sahni2015widespread}
\textbf{\color{eLifeMediumGrey} Sahni N}, Yi S, Taipale M, Bass JIF,
  Coulombe-Huntington J, Yang F, Peng J, Weile J, Karras GI, Wang Y, et~al.
\newblock Widespread macromolecular interaction perturbations in human genetic
  disorders.
\newblock Cell.  2015; 161(3):647--660.

\bibitem[{Salvi et~al.(2016)Salvi, Nicola and Abyzov, Anton and Blackledge,
  Martin}]{salvi2016multi}
\textbf{\color{eLifeMediumGrey} Salvi N}, Abyzov A, Blackledge M.
\newblock Multi-timescale dynamics in intrinsically disordered proteins from
  NMR relaxation and molecular simulation.
\newblock The journal of physical chemistry letters.  2016; 7(13):2483--2489.

\bibitem[{Sanborn et~al.(2021)Sanborn, Adrian L and Yeh, Benjamin T and
  Feigerle, Jordan T and Hao, Cynthia V and Townshend, Raphael JL and
  Lieberman-Aiden, Erez and Dror, Ron O and Kornberg, Roger
  D}]{sanborn2021simple}
\textbf{\color{eLifeMediumGrey} Sanborn AL}, Yeh BT, Feigerle JT, Hao CV,
  Townshend RJ, Lieberman-Aiden E, Dror RO, Kornberg RD.
\newblock Simple biochemical features underlie transcriptional activation
  domain diversity and dynamic, fuzzy binding to Mediator.
\newblock Elife.  2021; 10:e68068.

\bibitem[{Sawle and Ghosh(2015)Sawle, Lucas and Ghosh,
  Kingshuk}]{sawle2015theoretical}
\textbf{\color{eLifeMediumGrey} Sawle L}, Ghosh K.
\newblock A theoretical method to compute sequence dependent configurational
  properties in charged polymers and proteins.
\newblock The Journal of chemical physics.  2015; 143(8):08B615\_1.

\bibitem[{Schuler et~al.(2020)Schuler, Benjamin and Borgia, Alessandro and
  Borgia, Madeleine B and Heidarsson, P{\'e}tur O and Holmstrom, Erik D and
  Nettels, Daniel and Sottini, Andrea}]{schuler2020binding}
\textbf{\color{eLifeMediumGrey} Schuler B}, Borgia A, Borgia MB, Heidarsson PO,
  Holmstrom ED, Nettels D, Sottini A.
\newblock Binding without folding--the biomolecular function of disordered
  polyelectrolyte complexes.
\newblock Current opinion in structural biology.  2020; 60:66--76.

\bibitem[{Senior et~al.(2020)Senior, Andrew W and Evans, Richard and Jumper,
  John and Kirkpatrick, James and Sifre, Laurent and Green, Tim and Qin,
  Chongli and {\v{Z}}{\'\i}dek, Augustin and Nelson, Alexander WR and
  Bridgland, Alex and others}]{senior2020improved}
\textbf{\color{eLifeMediumGrey} Senior AW}, Evans R, Jumper J, Kirkpatrick J,
  Sifre L, Green T, Qin C, {\v{Z}}{\'\i}dek A, Nelson AW, Bridgland A, et~al.
\newblock Improved protein structure prediction using potentials from deep
  learning.
\newblock Nature.  2020; 577(7792):706--710.

\bibitem[{Seuma et~al.(2021)Seuma, Mireia and Faure, Andre J and Badia, Marta
  and Lehner, Ben and Bolognesi, Benedetta}]{seuma2021genetic}
\textbf{\color{eLifeMediumGrey} Seuma M}, Faure AJ, Badia M, Lehner B,
  Bolognesi B.
\newblock The genetic landscape for amyloid beta fibril nucleation accurately
  discriminates familial Alzheimer’s disease mutations.
\newblock Elife.  2021; 10:e63364.

\bibitem[{Shen and Bax(2007)Shen, Yang and Bax, Ad}]{shen2007protein}
\textbf{\color{eLifeMediumGrey} Shen Y}, Bax A.
\newblock Protein backbone chemical shifts predicted from searching a database
  for torsion angle and sequence homology.
\newblock Journal of biomolecular NMR.  2007; 38(4):289--302.

\bibitem[{Shen and Bax(2010)Shen, Yang and Bax, Ad}]{shen2010sparta+}
\textbf{\color{eLifeMediumGrey} Shen Y}, Bax A.
\newblock SPARTA+: a modest improvement in empirical NMR chemical shift
  prediction by means of an artificial neural network.
\newblock Journal of biomolecular NMR.  2010; 48(1):13--22.

\bibitem[{Shin and Brangwynne(2017)Shin, Yongdae and Brangwynne, Clifford
  P}]{shin2017liquid}
\textbf{\color{eLifeMediumGrey} Shin Y}, Brangwynne CP.
\newblock Liquid phase condensation in cell physiology and disease.
\newblock Science.  2017; 357(6357).

\bibitem[{Shlyueva et~al.(2014)Shlyueva, Daria and Stampfel, Gerald and Stark,
  Alexander}]{shlyueva2014transcriptional}
\textbf{\color{eLifeMediumGrey} Shlyueva D}, Stampfel G, Stark A.
\newblock Transcriptional enhancers: from properties to genome-wide
  predictions.
\newblock Nature Reviews Genetics.  2014; 15(4):272--286.

\bibitem[{Sigler(1988)Sigler, Paul B}]{sigler1988acid}
\textbf{\color{eLifeMediumGrey} Sigler PB}.
\newblock Acid blobs and negative noodles.
\newblock Nature.  1988; 333(6170):210--212.

\bibitem[{Skerker et~al.(2008)Skerker, Jeffrey M and Perchuk, Barrett S and
  Siryaporn, Albert and Lubin, Emma A and Ashenberg, Orr and Goulian, Mark and
  Laub, Michael T}]{skerker2008rewiring}
\textbf{\color{eLifeMediumGrey} Skerker JM}, Perchuk BS, Siryaporn A, Lubin EA,
  Ashenberg O, Goulian M, Laub MT.
\newblock Rewiring the specificity of two-component signal transduction
  systems.
\newblock Cell.  2008; 133(6):1043--1054.

\bibitem[{S{\o}rensen and Kjaergaard(2019)S{\o}rensen, Charlotte S and
  Kjaergaard, Magnus}]{sorensen2019effective}
\textbf{\color{eLifeMediumGrey} S{\o}rensen CS}, Kjaergaard M.
\newblock Effective concentrations enforced by intrinsically disordered linkers
  are governed by polymer physics.
\newblock Proceedings of the National Academy of Sciences.  2019;
  116(46):23124--23131.

\bibitem[{Sottini et~al.(2020)Sottini, Andrea and Borgia, Alessandro and
  Borgia, Madeleine B and Bugge, Katrine and Nettels, Daniel and Chowdhury,
  Aritra and Heidarsson, P{\'e}tur O and Zosel, Franziska and Best, Robert B
  and Kragelund, Birthe B and others}]{sottini2020polyelectrolyte}
\textbf{\color{eLifeMediumGrey} Sottini A}, Borgia A, Borgia MB, Bugge K,
  Nettels D, Chowdhury A, Heidarsson PO, Zosel F, Best RB, Kragelund BB, et~al.
\newblock Polyelectrolyte interactions enable rapid association and
  dissociation in high-affinity disordered protein complexes.
\newblock Nature communications.  2020; 11(1):1--14.

\bibitem[{Staller et~al.(2018)Staller, Max V and Holehouse, Alex S and
  Swain-Lenz, Devjanee and Das, Rahul K and Pappu, Rohit V and Cohen, Barak
  A}]{staller2018high}
\textbf{\color{eLifeMediumGrey} Staller MV}, Holehouse AS, Swain-Lenz D, Das
  RK, Pappu RV, Cohen BA.
\newblock A high-throughput mutational scan of an intrinsically disordered
  acidic transcriptional activation domain.
\newblock Cell systems.  2018; 6(4):444--455.

\bibitem[{Staller et~al.(2021)Staller, Max V and Ramirez, Eddie and Holehouse,
  Alex S and Pappu, Rohit V and Cohen, Barak A}]{staller2021design}
\textbf{\color{eLifeMediumGrey} Staller MV}, Ramirez E, Holehouse AS, Pappu RV,
  Cohen BA.
\newblock Design principles of acidic transcriptional activation domains.
\newblock bioRxiv.  2021; p. 2020--10.

\bibitem[{Starita et~al.(2017)Starita, Lea M and Ahituv, Nadav and Dunham,
  Maitreya J and Kitzman, Jacob O and Roth, Frederick P and Seelig, Georg and
  Shendure, Jay and Fowler, Douglas M}]{starita2017variant}
\textbf{\color{eLifeMediumGrey} Starita LM}, Ahituv N, Dunham MJ, Kitzman JO,
  Roth FP, Seelig G, Shendure J, Fowler DM.
\newblock Variant interpretation: functional assays to the rescue.
\newblock The American Journal of Human Genetics.  2017; 101(3):315--325.

\bibitem[{Statt et~al.(2020)Statt, Antonia and Casademunt, Helena and
  Brangwynne, Clifford P and Panagiotopoulos, Athanassios Z}]{statt2020model}
\textbf{\color{eLifeMediumGrey} Statt A}, Casademunt H, Brangwynne CP,
  Panagiotopoulos AZ.
\newblock Model for disordered proteins with strongly sequence-dependent liquid
  phase behavior.
\newblock The Journal of chemical physics.  2020; 152(7):075101.

\bibitem[{Stefl et~al.(2013)Stefl, Shannon and Nishi, Hafumi and Petukh,
  Marharyta and Panchenko, Anna R and Alexov, Emil}]{stefl2013molecular}
\textbf{\color{eLifeMediumGrey} Stefl S}, Nishi H, Petukh M, Panchenko AR,
  Alexov E.
\newblock Molecular mechanisms of disease-causing missense mutations.
\newblock Journal of molecular biology.  2013; 425(21):3919--3936.

\bibitem[{Stein and Aloy(2008)Stein, Amelie and Aloy,
  Patrick}]{stein2008contextual}
\textbf{\color{eLifeMediumGrey} Stein A}, Aloy P.
\newblock Contextual specificity in peptide-mediated protein interactions.
\newblock PloS one.  2008; 3(7):e2524.

\bibitem[{Stein et~al.(2019)Stein, Amelie and Fowler, Douglas M and
  Hartmann-Petersen, Rasmus and Lindorff-Larsen,
  Kresten}]{stein2019biophysical}
\textbf{\color{eLifeMediumGrey} Stein A}, Fowler DM, Hartmann-Petersen R,
  Lindorff-Larsen K.
\newblock Biophysical and mechanistic models for disease-causing protein
  variants.
\newblock Trends in biochemical sciences.  2019; 44(7):575--588.

\bibitem[{Sugase et~al.(2007)Sugase, Kenji and Dyson, H Jane and Wright, Peter
  E}]{sugase2007mechanism}
\textbf{\color{eLifeMediumGrey} Sugase K}, Dyson HJ, Wright PE.
\newblock Mechanism of coupled folding and binding of an intrinsically
  disordered protein.
\newblock Nature.  2007; 447(7147):1021--1025.

\bibitem[{Sundell et~al.(2018)Sundell, Gustav N and Arnold, Roland and Ali,
  Muhammad and Naksukpaiboon, Piangfan and Orts, Julien and G{\"u}ntert, Peter
  and Chi, Celestine N and Ivarsson, Ylva}]{sundell2018proteome}
\textbf{\color{eLifeMediumGrey} Sundell GN}, Arnold R, Ali M, Naksukpaiboon P,
  Orts J, G{\"u}ntert P, Chi CN, Ivarsson Y.
\newblock Proteome-wide analysis of phospho-regulated PDZ domain interactions.
\newblock Molecular systems biology.  2018; 14(8):e8129.

\bibitem[{Svergun et~al.(1995)Svergun, Dmitri and Barberato, Claudio and Koch,
  Michel HJ}]{svergun1995crysol}
\textbf{\color{eLifeMediumGrey} Svergun D}, Barberato C, Koch MH.
\newblock CRYSOL--a program to evaluate X-ray solution scattering of biological
  macromolecules from atomic coordinates.
\newblock Journal of applied crystallography.  1995; 28(6):768--773.

\bibitem[{Swanson et~al.(2004)Swanson, Kurt A and Knoepfler, Paul S and Huang,
  Kai and Kang, Richard S and Cowley, Shaun M and Laherty, Carol D and
  Eisenman, Robert N and Radhakrishnan, Ishwar}]{swanson2004hbp1}
\textbf{\color{eLifeMediumGrey} Swanson KA}, Knoepfler PS, Huang K, Kang RS,
  Cowley SM, Laherty CD, Eisenman RN, Radhakrishnan I.
\newblock HBP1 and Mad1 repressors bind the Sin3 corepressor PAH2 domain with
  opposite helical orientations.
\newblock Nature structural \& molecular biology.  2004; 11(8):738--746.

\bibitem[{Taylor and Hatrick(1994)Taylor, William R and Hatrick,
  Kerr}]{taylor1994compensating}
\textbf{\color{eLifeMediumGrey} Taylor WR}, Hatrick K.
\newblock Compensating changes in protein multiple sequence alignments.
\newblock Protein Engineering, Design and Selection.  1994; 7(3):341--348.

\bibitem[{Teilum et~al.(2021)Teilum, Kaare and Olsen, Johan G and Kragelund,
  Birthe B}]{teilum2021new}
\textbf{\color{eLifeMediumGrey} Teilum K}, Olsen JG, Kragelund BB.
\newblock On the specificity of protein-protein interactions in the context of
  disorder.
\newblock Biochemical Journal.  2021; in press.

\bibitem[{Tian et~al.(2015)Tian, Pengfei and Boomsma, Wouter and Wang, Yong and
  Otzen, Daniel E and Jensen, Mogens H and Lindorff-Larsen,
  Kresten}]{tian2015structure}
\textbf{\color{eLifeMediumGrey} Tian P}, Boomsma W, Wang Y, Otzen DE, Jensen
  MH, Lindorff-Larsen K.
\newblock Structure of a functional amyloid protein subunit computed using
  sequence variation.
\newblock Journal of the American Chemical Society.  2015; 137(1):22--25.

\bibitem[{Tiberti et~al.(2015)Tiberti, Matteo and Papaleo, Elena and Bengtsen,
  Tone and Boomsma, Wouter and Lindorff-Larsen, Kresten}]{tiberti2015encore}
\textbf{\color{eLifeMediumGrey} Tiberti M}, Papaleo E, Bengtsen T, Boomsma W,
  Lindorff-Larsen K.
\newblock ENCORE: software for quantitative ensemble comparison.
\newblock PLoS Comput Biol.  2015; 11(10):e1004415.

\bibitem[{Tillu et~al.(2021)Tillu, Vikas A and Rae, James and Gao, Ya and
  Ariotti, Nicholas and Floetenmeyer, Matthias and Kovtun, Oleksiy and McMahon,
  Kerrie-Ann and Chaudhary, Natasha and Parton, Robert G and Collins, Brett
  M}]{tillu2021cavin1}
\textbf{\color{eLifeMediumGrey} Tillu VA}, Rae J, Gao Y, Ariotti N,
  Floetenmeyer M, Kovtun O, McMahon KA, Chaudhary N, Parton RG, Collins BM.
\newblock Cavin1 intrinsically disordered domains are essential for fuzzy
  electrostatic interactions and caveola formation.
\newblock Nature communications.  2021; 12(1):1--18.

\bibitem[{Tompa et~al.(2014)Tompa, Peter and Davey, Norman E and Gibson, Toby J
  and Babu, M Madan}]{tompa2014million}
\textbf{\color{eLifeMediumGrey} Tompa P}, Davey NE, Gibson TJ, Babu MM.
\newblock A million peptide motifs for the molecular biologist.
\newblock Molecular cell.  2014; 55(2):161--169.

\bibitem[{Torrisi et~al.(2020)Torrisi, Mirko and Pollastri, Gianluca and Le,
  Quan}]{torrisi2020deep}
\textbf{\color{eLifeMediumGrey} Torrisi M}, Pollastri G, Le Q.
\newblock Deep learning methods in protein structure prediction.
\newblock Computational and Structural Biotechnology Journal.  2020; .

\bibitem[{Toth-Petroczy et~al.(2016)Toth-Petroczy, Agnes and Palmedo, Perry and
  Ingraham, John and Hopf, Thomas A and Berger, Bonnie and Sander, Chris and
  Marks, Debora S}]{toth2016structured}
\textbf{\color{eLifeMediumGrey} Toth-Petroczy A}, Palmedo P, Ingraham J, Hopf
  TA, Berger B, Sander C, Marks DS.
\newblock Structured states of disordered proteins from genomic sequences.
\newblock Cell.  2016; 167(1):158--170.

\bibitem[{Tsang et~al.(2020)Tsang, Brian and Priti{\v{s}}anac, Iva and Scherer,
  Stephen W and Moses, Alan M and Forman-Kay, Julie D}]{tsang2020phase}
\textbf{\color{eLifeMediumGrey} Tsang B}, Priti{\v{s}}anac I, Scherer SW, Moses
  AM, Forman-Kay JD.
\newblock Phase Separation as a Missing Mechanism for Interpretation of Disease
  Mutations.
\newblock Cell.  2020; 183(7):1742--1756.

\bibitem[{Tycko et~al.(2020)Tycko, Josh and DelRosso, Nicole and Hess, Gaelen T
  and Banerjee, Abhimanyu and Mukund, Aditya and Van, Mike V and Ego, Braeden K
  and Yao, David and Spees, Kaitlyn and Suzuki, Peter and
  others}]{tycko2020high}
\textbf{\color{eLifeMediumGrey} Tycko J}, DelRosso N, Hess GT, Banerjee A,
  Mukund A, Van MV, Ego BK, Yao D, Spees K, Suzuki P, et~al.
\newblock High-throughput discovery and characterization of human
  transcriptional effectors.
\newblock Cell.  2020; .

\bibitem[{Uversky(2013)Uversky, Vladimir N}]{uversky2013most}
\textbf{\color{eLifeMediumGrey} Uversky VN}.
\newblock The most important thing is the tail: multitudinous functionalities
  of intrinsically disordered protein termini.
\newblock FEBS letters.  2013; 587(13):1891--1901.

\bibitem[{Uversky(2015)Uversky, Vladimir N}]{uversky2015intrinsically}
\textbf{\color{eLifeMediumGrey} Uversky VN}.
\newblock Intrinsically disordered proteins and their (disordered) proteomes in
  neurodegenerative disorders.
\newblock Frontiers in aging neuroscience.  2015; 7:18.

\bibitem[{Uversky et~al.(2009)Uversky, Vladimir N and Oldfield, Christopher J
  and Midic, Uros and Xie, Hongbo and Xue, Bin and Vucetic, Slobodan and
  Iakoucheva, Lilia M and Obradovic, Zoran and Dunker, A
  Keith}]{uversky2009unfoldomics}
\textbf{\color{eLifeMediumGrey} Uversky VN}, Oldfield CJ, Midic U, Xie H, Xue
  B, Vucetic S, Iakoucheva LM, Obradovic Z, Dunker AK.
\newblock Unfoldomics of human diseases: linking protein intrinsic disorder
  with diseases.
\newblock BMC genomics.  2009; 10(1):1--17.

\bibitem[{Vacic et~al.(2012)Vacic, Vladimir and Markwick, Phineus RL and
  Oldfield, Christopher J and Zhao, Xiaoyue and Haynes, Chad and Uversky,
  Vladimir N and Iakoucheva, Lilia M}]{vacic2012disease}
\textbf{\color{eLifeMediumGrey} Vacic V}, Markwick PR, Oldfield CJ, Zhao X,
  Haynes C, Uversky VN, Iakoucheva LM.
\newblock Disease-associated mutations disrupt functionally important regions
  of intrinsic protein disorder.
\newblock PLoS Comput Biol.  2012; 8(10):e1002709.

\bibitem[{Van~Roey et~al.(2014)Van Roey, Kim and Uyar, Bora and Weatheritt,
  Robert J and Dinkel, Holger and Seiler, Markus and Budd, Aidan and Gibson,
  Toby J and Davey, Norman E}]{van2014short}
\textbf{\color{eLifeMediumGrey} Van~Roey K}, Uyar B, Weatheritt RJ, Dinkel H,
  Seiler M, Budd A, Gibson TJ, Davey NE.
\newblock Short linear motifs: ubiquitous and functionally diverse protein
  interaction modules directing cell regulation.
\newblock Chemical reviews.  2014; 114(13):6733--6778.

\bibitem[{Vernon et~al.(2018)Vernon, Robert McCoy and Chong, Paul Andrew and
  Tsang, Brian and Kim, Tae Hun and Bah, Alaji and Farber, Patrick and Lin,
  Hong and Forman-Kay, Julie Deborah}]{vernon2018pi}
\textbf{\color{eLifeMediumGrey} Vernon RM}, Chong PA, Tsang B, Kim TH, Bah A,
  Farber P, Lin H, Forman-Kay JD.
\newblock Pi-Pi contacts are an overlooked protein feature relevant to phase
  separation.
\newblock elife.  2018; 7:e31486.

\bibitem[{Wallweber et~al.(2014)Wallweber, Heidi JA and Tam, Christine and
  Franke, Yvonne and Starovasnik, Melissa A and Lupardus, Patrick
  J}]{wallweber2014structural}
\textbf{\color{eLifeMediumGrey} Wallweber HJ}, Tam C, Franke Y, Starovasnik MA,
  Lupardus PJ.
\newblock Structural basis of recognition of interferon-$\alpha$ receptor by
  tyrosine kinase 2.
\newblock Nature structural \& molecular biology.  2014; 21(5):443.

\bibitem[{Wang et~al.(2018)Wang, Jie and Choi, Jeong-Mo and Holehouse, Alex S
  and Lee, Hyun O and Zhang, Xiaojie and Jahnel, Marcus and Maharana,
  Shovamayee and Lemaitre, R{\'e}gis and Pozniakovsky, Andrei and Drechsel,
  David and others}]{wang2018molecular}
\textbf{\color{eLifeMediumGrey} Wang J}, Choi JM, Holehouse AS, Lee HO, Zhang
  X, Jahnel M, Maharana S, Lemaitre R, Pozniakovsky A, Drechsel D, et~al.
\newblock A molecular grammar governing the driving forces for phase separation
  of prion-like RNA binding proteins.
\newblock Cell.  2018; 174(3):688--699.

\bibitem[{Wang et~al.(2014)Wang, Lee-Ping and Martinez, Todd J and Pande, Vijay
  S}]{wang2014building}
\textbf{\color{eLifeMediumGrey} Wang LP}, Martinez TJ, Pande VS.
\newblock Building force fields: An automatic, systematic, and reproducible
  approach.
\newblock The journal of physical chemistry letters.  2014; 5(11):1885--1891.

\bibitem[{Weigt et~al.(2009)Weigt, Martin and White, Robert A and Szurmant,
  Hendrik and Hoch, James A and Hwa, Terence}]{weigt2009identification}
\textbf{\color{eLifeMediumGrey} Weigt M}, White RA, Szurmant H, Hoch JA, Hwa T.
\newblock Identification of direct residue contacts in protein--protein
  interaction by message passing.
\newblock Proceedings of the National Academy of Sciences.  2009;
  106(1):67--72.

\bibitem[{Wheeler et~al.(2020)Wheeler, Lucas C and Perkins, Arden and Wong,
  Caitlyn E and Harms, Michael J}]{wheeler2020learning}
\textbf{\color{eLifeMediumGrey} Wheeler LC}, Perkins A, Wong CE, Harms MJ.
\newblock Learning peptide recognition rules for a low-specificity protein.
\newblock Protein Science.  2020; 29(11):2259--2273.

\bibitem[{Wigington et~al.(2020)Wigington, Callie P and Roy, Jagoree and Damle,
  Nikhil P and Yadav, Vikash K and Blikstad, Cecilia and Resch, Eduard and
  Wong, Cassandra J and Mackay, Douglas R and Wang, Jennifer T and Krystkowiak,
  Izabella and others}]{wigington2020systematic}
\textbf{\color{eLifeMediumGrey} Wigington CP}, Roy J, Damle NP, Yadav VK,
  Blikstad C, Resch E, Wong CJ, Mackay DR, Wang JT, Krystkowiak I, et~al.
\newblock Systematic discovery of Short Linear Motifs decodes calcineurin
  phosphatase signaling.
\newblock Molecular Cell.  2020; 79(2):342--358.

\bibitem[{Wong et~al.(2020)Wong, Eric TC and So, Victor and Guron, Mike and
  Kuechler, Erich R and Malhis, Nawar and Bui, Jennifer M and Gsponer,
  J{\"o}rg}]{wong2020protein}
\textbf{\color{eLifeMediumGrey} Wong ET}, So V, Guron M, Kuechler ER, Malhis N,
  Bui JM, Gsponer J.
\newblock Protein--protein interactions mediated by intrinsically disordered
  protein regions are enriched in missense mutations.
\newblock Biomolecules.  2020; 10(8):1097.

\bibitem[{Worswick et~al.(2018)Worswick, Steven G and Spencer, James A and
  Jeschke, Gunnar and Kuprov, Ilya}]{worswick2018deep}
\textbf{\color{eLifeMediumGrey} Worswick SG}, Spencer JA, Jeschke G, Kuprov I.
\newblock Deep neural network processing of DEER data.
\newblock Science advances.  2018; 4(8):eaat5218.

\bibitem[{Xu(2019)Xu, Jinbo}]{xu2019distance}
\textbf{\color{eLifeMediumGrey} Xu J}.
\newblock Distance-based protein folding powered by deep learning.
\newblock Proceedings of the National Academy of Sciences.  2019;
  116(34):16856--16865.

\bibitem[{Xu and Case(2001)Xu, Xiao-Ping and Case, David A}]{xu2001automated}
\textbf{\color{eLifeMediumGrey} Xu XP}, Case DA.
\newblock {Automated prediction of 15N, 13C$\alpha$, 13C$\beta$ and 13C'
  chemical shifts in proteins using a density functional database}.
\newblock Journal of biomolecular NMR.  2001; 21(4):321--333.

\bibitem[{Yang et~al.(2021)Yang, Huan and Xiong, Zhaoping and Zonta,
  Francesco}]{yang2021construction}
\textbf{\color{eLifeMediumGrey} Yang H}, Xiong Z, Zonta F.
\newblock Construction of a neural network energy function for protein physics.
\newblock bioRxiv.  2021; .

\bibitem[{Yang et~al.(2020{\natexlab{a}})Yang, Jianyi and Anishchenko, Ivan and
  Park, Hahnbeom and Peng, Zhenling and Ovchinnikov, Sergey and Baker,
  David}]{yang2020improved}
\textbf{\color{eLifeMediumGrey} Yang J}, Anishchenko I, Park H, Peng Z,
  Ovchinnikov S, Baker D.
\newblock Improved protein structure prediction using predicted interresidue
  orientations.
\newblock Proceedings of the National Academy of Sciences.  2020;
  117(3):1496--1503.

\bibitem[{Yang et~al.(2020{\natexlab{b}})Yang, Ziyue and Chakraborty, Maghesree
  and White, Andrew D}]{yang2020predicting}
\textbf{\color{eLifeMediumGrey} Yang Z}, Chakraborty M, White AD.
\newblock Predicting Chemical Shifts with Graph Neural Networks.
\newblock bioRxiv.  2020; .

\bibitem[{Ye et~al.(2020)Ye, Sheng and Zhong, Kai and Zhang, Jinxiao and Hu,
  Wei and Hirst, Jonathan D and Zhang, Guozhen and Mukamel, Shaul and Jiang,
  Jun}]{ye2020machine}
\textbf{\color{eLifeMediumGrey} Ye S}, Zhong K, Zhang J, Hu W, Hirst JD, Zhang
  G, Mukamel S, Jiang J.
\newblock A Machine Learning Protocol for Predicting Protein Infrared Spectra.
\newblock Journal of the American Chemical Society.  2020;
  142(45):19071--19077.

\bibitem[{You et~al.(2020)You, Kaiqiang and Huang, Qi and Yu, Chunyu and Shen,
  Boyan and Sevilla, Cristoffer and Shi, Minglei and Hermjakob, Henning and
  Chen, Yang and Li, Tingting}]{you2020phasepdb}
\textbf{\color{eLifeMediumGrey} You K}, Huang Q, Yu C, Shen B, Sevilla C, Shi
  M, Hermjakob H, Chen Y, Li T.
\newblock PhaSepDB: a database of liquid--liquid phase separation related
  proteins.
\newblock Nucleic acids research.  2020; 48(D1):D354--D359.

\bibitem[{Zarin et~al.(2019)Zarin, Taraneh and Strome, Bob and Ba, Alex N
  Nguyen and Alberti, Simon and Forman-Kay, Julie D and Moses, Alan
  M}]{zarin2019proteome}
\textbf{\color{eLifeMediumGrey} Zarin T}, Strome B, Ba ANN, Alberti S,
  Forman-Kay JD, Moses AM.
\newblock Proteome-wide signatures of function in highly diverged intrinsically
  disordered regions.
\newblock Elife.  2019; 8:e46883.

\bibitem[{Zarin et~al.(2021)Zarin, Taraneh and Strome, Bob and Peng, Gang and
  Priti{\v{s}}anac, Iva and Forman-Kay, Julie D and Moses, Alan
  M}]{zarin2021identifying}
\textbf{\color{eLifeMediumGrey} Zarin T}, Strome B, Peng G, Priti{\v{s}}anac I,
  Forman-Kay JD, Moses AM.
\newblock Identifying molecular features that are associated with biological
  function of intrinsically disordered protein regions.
\newblock Elife.  2021; 10:e60220.

\bibitem[{Zeke et~al.(2015)Zeke, Andr{\'a}s and Bastys, Tomas and Alexa, Anita
  and Garai, {\'A}gnes and M{\'e}sz{\'a}ros, B{\'a}lint and Kirsch, Kl{\'a}ra
  and Doszt{\'a}nyi, Zsuzsanna and Kalinina, Olga V and Rem{\'e}nyi,
  Attila}]{zeke2015systematic}
\textbf{\color{eLifeMediumGrey} Zeke A}, Bastys T, Alexa A, Garai {\'A},
  M{\'e}sz{\'a}ros B, Kirsch K, Doszt{\'a}nyi Z, Kalinina OV, Rem{\'e}nyi A.
\newblock Systematic discovery of linear binding motifs targeting an ancient
  protein interaction surface on MAP kinases.
\newblock Molecular systems biology.  2015; 11(11):837.

\bibitem[{Zheng et~al.(2019)Zheng, Wei and Li, Yang and Zhang, Chengxin and
  Pearce, Robin and Mortuza, SM and Zhang, Yang}]{zheng2019deep}
\textbf{\color{eLifeMediumGrey} Zheng W}, Li Y, Zhang C, Pearce R, Mortuza S,
  Zhang Y.
\newblock Deep-learning contact-map guided protein structure prediction in
  CASP13.
\newblock Proteins: Structure, Function, and Bioinformatics.  2019;
  87(12):1149--1164.

\bibitem[{Zheng and Best(2018)Zheng, Wenwei and Best, Robert
  B}]{zheng2018extended}
\textbf{\color{eLifeMediumGrey} Zheng W}, Best RB.
\newblock An extended Guinier analysis for intrinsically disordered proteins.
\newblock Journal of molecular biology.  2018; 430(16):2540--2553.

\bibitem[{Zheng et~al.(2020)Zheng, Wenwei and Dignon, Gregory and Brown,
  Matthew and Kim, Young C and Mittal, Jeetain}]{zheng2020hydropathy}
\textbf{\color{eLifeMediumGrey} Zheng W}, Dignon G, Brown M, Kim YC, Mittal J.
\newblock Hydropathy patterning complements charge patterning to describe
  conformational preferences of disordered proteins.
\newblock The journal of physical chemistry letters.  2020; 11(9):3408--3415.

\bibitem[{Zhou et~al.(2020)Zhou, Jing-Bo and Xiong, Yao and An, Ke and Ye,
  Zhi-Qiang and Wu, Yun-Dong}]{zhou2020idrmutpred}
\textbf{\color{eLifeMediumGrey} Zhou JB}, Xiong Y, An K, Ye ZQ, Wu YD.
\newblock IDRMutPred: predicting disease-associated germline nonsynonymous
  single nucleotide variants (nsSNVs) in intrinsically disordered regions.
\newblock Bioinformatics.  2020; 36(20):4977--4983.

\end{thebibliography}

\end{document}